\documentclass[fleqn,usenatbib]{mnras}

\usepackage{newtxtext,newtxmath}
\usepackage[T1]{fontenc}
\usepackage{soul}

\DeclareRobustCommand{\VAN}[3]{#2}
\let\VANthebibliography\thebibliography
\def\thebibliography{\DeclareRobustCommand{\VAN}[3]{##3}\VANthebibliography}

\usepackage{graphicx}	% Including figure files
\usepackage{amsmath}	% Advanced maths commands
\usepackage{orcidlink}

\makeatletter
\renewcommand{\fixfootnotes}{%
  \renewcommand{\thefootnote}{\arabic{footnote}}%
  \renewcommand{\@makefnmark}{\textsuperscript{[\@thefnmark]}}%
  \long\def\@makefntext##1{%
    \parindent 1em\noindent
    \textsuperscript{[\@thefnmark]}\hspace{4pt}##1%
  }%
  \setcounter{footnote}{0}%
}
\makeatother

\newcommand{\revision}[1]{\textcolor{black}{#1}}

\title[Viscous vs MHD-driven evolution of discs]{Using EUV driven external photoevaporation to test viscous evolution of protoplanetary discs}

\author[Ballabio \& Owen]{
Giulia Ballabio$^{\orcidlink{0000-0002-4687-2133}1}$\thanks{E-mail: g.ballabio@imperial.ac.uk}
and James E. Owen$^{\orcidlink{0000-0002-4856-7837}1,2}$
\\
$^{1}$Imperial Astrophysics, Imperial College London, Blackett Laboratory, Prince Consort Road, London SW7 2AZ, UK\\
$^{2}$Department of Earth, Planetary, and Space Sciences, University of California, Los Angeles, CA 90095, USA
}

\date{Accepted 2026 February 13. Received 2026 February 11; in original form 2025 August 29}

\pubyear{2024}

\begin{document}
\label{firstpage}
\pagerange{\pageref{firstpage}--\pageref{lastpage}}
\maketitle

\begin{abstract}
Protoplanetary discs are thought to evolve either through angular momentum transport driven by viscous processes or through angular momentum removal induced by magnetohydrodynamic (MHD) winds. One proposed method to distinguish between these two evolutionary pathways is by comparing mass accretion rates and disc sizes, but observational constraints complicate this distinction. In this study, we investigate how extreme ultraviolet (EUV) external photoevaporation affects the evolution of protoplanetary discs, particularly in environments such as the Orion Nebula Cluster. Using a combination of analytical derivations and 1D numerical simulations, we explore the impact of externally induced mass-loss on disc structure and accretion dynamics. We demonstrate that, in the viscous scenario, there exists a clear, near one-to-one correlation between the mass-loss rate due to external photoevaporative outflows and the mass accretion rate onto the central star. In contrast, MHD wind-driven discs do not exhibit such trend, leading to a distinct evolutionary path. External photoevaporative mass-loss rates and mass accretion rates can both be accurately measured for a population of discs, without a strong model dependence. Thus, our findings provide a robust observational test to distinguish between viscous and MHD wind-driven disc evolution, offering a new approach to constraining angular momentum transport mechanisms in protoplanetary discs. \revision{Applying this diagnostic observationally requires joint measurements of $\dot{M}_{\rm acc}$ and $\dot{M}_{\rm pe}$ for the same objects, which are currently scarce in bright H\,\textsc{ii} regions due to contamination and sensitivity limitations.}
\end{abstract}

\begin{keywords}
planets and satellites: atmospheres -- planets and satellites: physical evolution -- radiative transfer
\end{keywords}

%%%%%%%%%%%%%%%%%%%%%%%%%%%%%%%%%%%%%%%%%%%%%%%%%%

%%%%%%%%%%%%%%%%% BODY OF PAPER %%%%%%%%%%%%%%%%%%

\section{Introduction} \label{sec:intro}

Planets form and evolve within protoplanetary discs of gas and dust \citep[e.g.,][]{2018A&A...617A..44K, 2018ApJ...860L..13P,Armitage2018,2019Natur.574..378T}. Over a period of approximately 1–10 million years, these discs disperse, driven by processes such as accretion onto the central (proto-)star and mass-loss driven by thermal or magnetic forces \citep[e.g.,][]{2017RSOS....470114E, 2022AAS...24041302P, 2023ASPC..534..465L, 2023ASPC..534..539M}. However, key aspects of planet formation during the disc phase remain poorly understood, particularly because we do not have a complete understanding of how angular momentum is transported within these discs. This transport plays a critical role in both the growth and migration of planets \citep[e.g.,][]{2014prpl.conf..667B, 2023ASPC..534..465L}, as well as dust evolution \citep[e.g.,][]{2014prpl.conf..339T}. However, it is currently unclear whether angular momentum is redistributed through viscous transport or removed by magnetohydrodynamic (MHD) winds. 
% While molecular viscosity is insufficient to explain observed accretion rates, other mechanisms like magnetorotational instability or gravitational torques may play a role. 
The widely used $\alpha$-viscosity models \citep{1973A&A....24..337S,1974MNRAS.168..603L} are often used to describe the evolution of the disc, but their appropriateness, and intensity of turbulence within the discs remain undetermined \revision{\citep[e.g.,][]{2023ASPC..534..501M, 2023NewAR..9601674R}}. 

% Observations of discs in different star-forming regions suggest conflicting evidence for viscous expansion, with some studies indicating extended discs and others pointing to more compact ones. However, disc radii may not provide reliable evidence due to decoupling between gas and dust or opacity effects. Gas disc radii, although promising, are still challenging to measure due to limited CO detections. 

In theory, measuring disc radii should the be an ideal method to differentiate between viscously driven and MHD-wind driven disc evolution \revision{\citep[][]{2018ApJ...864..168N, 2020A&A...640A...5T, 2022ApJ...926...61T, 2022MNRAS.514.1088Z}}. However, a significant challenge is the combined effect radial drift of dust grains and their opacity has on the measured radii of the dust disc \citep{2019MNRAS.486L..63R, 2019MNRAS.486.4829R}. Comparative studies of viscous and MHD models \citep{2022MNRAS.514.1088Z} confirm that dust radii are expected to expand only in the case of a viscous scenario, but observing this expansion requires very deep and costly observations. Consequently, the focus has shifted to the use of gas tracers to determine disc radii \citep[e.g.,][]{2023ASPC..534..501M}. This approach presents its own challenges because the gas is much harder to detect and, therefore, this approach is limited by the number of resolved CO detections. Therefore, it has proved difficult to preform these analysis on a large enough population of discs to yield conclusive answers. 

Alternatively, distinguishing between viscous evolution and magnetohydrodynamic wind-driven evolution can be approached by measuring mass accretion rates and global disc masses. In the viscous framework, a tight correlation is expected between the mass of the disc and the mass accretion rate, as both quantities decrease over time in a similar way \citep{hartmann1998, dullemond2006}. In contrast, 
no clear correlation between these quantities is predicted in the MHD wind-driven case, since the removal of angular momentum by MHD-winds alters the mass distribution and accretion process. 
Observationally testing these scenarios requires accurate measurements of both parameters, with disc masses being particularly challenging to determine \citep[e.g.,][]{2023ASPC..534..501M}. They are typically inferred from dust continuum emission, but uncertainties in dust opacity and gas-to-dust ratio complicate these estimates. Mass accretion rates are derived from ultraviolet and optical emission lines tracing accretion shocks as these measurements are typically much more robust, but variability and extinction introduce additional uncertainties \citep[e.g.,][]{2023ASPC..534..539M}. 

The key physics behind many of these tests is that in the viscous scenario the accretion rate onto the star must be linked to angular momentum transport at large radii. Hence, in a viscous disc, the accretion rate should correlate with the properties of the outer disc. In the MHD-wind scenario, no such causal link between the accretion close to the star and the properties of the outer disc is required. Therefore, these distinguishing methods can be thought of as null hypothesis tests demonstrating the existence of a correlation between the accretion rate and properties linked to the outer disc conditions. Although the accretion rate can be measured fairly robustly, all previous proposed properties related to the outer disc (size--both in gas and dust -- and mass) are uncertain, often by orders of magnitude. In this work, we propose an alternative property linked to the outer disc properties: the external photoevaporation rate. Under typical conditions in many star-forming regions the external photoevaporation rate can also be accurately measured. Therefore, in this work, we explore the expected correlations between accretion rates and external photoevaporation rates in both viscous and MHD-wind driven discs. 

\revision{\subsection{External photoevaporation}}
\label{intro_ext_pe}

Protoplanetary discs in dense star-forming regions are exposed to extreme ultraviolet (EUV; $h\nu > 13.6$eV) and far-ultraviolet (FUV; 6eV~$<h\nu<$13.6eV) radiation emitted primarily by nearby stars, such as 
% in the Trapezium cluster, particularly 
$\theta^1$C Ori in the case of Orion \citep{1994ApJ...428..654H, 1998ApJ...499..758J, 2000ApJ...539..258R, 2004ApJ...611..360A, 2016MNRAS.457.3593F, 2018MNRAS.481..452H}. 
EUV photons ionize the gas in the disc and heat it to temperatures around 10,000 K, whereas FUV photons primarily heat neutral gas in the photodissociation region (PDR) to about 100 to a few 1,000 K \citep[e.g.][]{1998ApJ...499..758J}. EUV photons are absorbed first, at the ionization front, while FUV photons penetrate deeper into the flow, affecting the neutral gas. When the heated gas gains enough thermal energy to overcome the gravitational energy, it escapes in the form of a hydrodynamic outflow. This occurs beyond a specific radius, the so-called gravitational radius, defined as $R_{\rm g} = GM_*/c_s^2$, where $M_*$ is the mass of the star and $c_s$ is the gas sound speed. The gravitational radius is approximately 9~au for a solar mass star, for EUV heated gas at 10$^{4}$~K. Although the photoevaporative outflow can be launched interior to the gravitational radius \citep[e.g.][]{Liffman2003}, the mass loss rates are exponentially suppressed \citep[e.g.][]{2004ApJ...611..360A,Owen2023}

Mass-loss rates strongly depend on whether the flow is dominated by EUV or FUV radiation, with stronger EUV radiation pushing the ionization front closer to the disc's surface, while strong FUV-driven outflows push the ionization front further away \citep[][]{1998ApJ...499..758J, 1999ApJ...515..669S}. This interplay between EUV and FUV irradiation is crucial to understand how external radiation influences the long-term evolution and destruction of protoplanetary discs. 
Recent studies have shown that external photoevaporation significantly impacts disc evolution, with mass-loss rates suggesting that many discs in the solar neighbourhood experience some photoevaporative depletion \revision{\citep[e.g.,][]{2022EPJP..137.1132W, 2023ASPC..534..539M, 2025A&A...695A..74A}}. Although extreme mass loss rates observed in regions like the Orion Nebula Cluster (ONC) probably represent short-lived phases \citep[e.g.,][]{1996AJ....111.1977M, 1999AJ....118.2350H, 2016arXiv160501773G, 2018MNRAS.475.5460H, 2019MNRAS.490.5678C, 2019MNRAS.490.5478W, 2020MNRAS.491..903W, 2020MNRAS.492.1279S, 2022MNRAS.512.3788Q}, they offer insight into how external photoevaporation truncates discs, affecting their evolution. The bright proplyds in the ONC were the first to be observed \citep{1994ApJ...436..194O, 1998AJ....115..263O, 1999AJ....118.2350H}, but recent surveys have discovered many more sources that undergo mass depletion in regions with lower UV fluxes \revision{\citep[e.g.,][]{2016ApJ...826L..15K, 2021MNRAS.501.3502H, 2024ApJ...960...49R}}. The ONC is still the star-forming region with the highest number of detected proplyds to date, numbering $\sim 150$ members \revision{\citep[][]{1993ApJ...410..696O, 1998ApJ...499..758J, 2008AJ....136.2136R}}. 
Mass-loss rates from proplyds have traditionally been inferred from observational analysis of emission lines, associated with ionized gas, such as H$\alpha$ \citep{henney1998}, [OIII] and [NII] \revision{\citep[e.g,][]{1999AJ....118.2350H, 2024A&A...687A..93A}}, which trace the outflowing material. 
More recently, EUV-driven mass-loss rates have also been measured using radio free-free emission and hydrogen recombination lines \citep[e.g.,][]{2023ApJ...954..127B, 2024ApJ...967..103B}. 
Measurements of the surface brightness and spatial distribution of these lines allow calculation of the electron density and velocity of the ionized flow, providing a robust method for deriving mass-loss rates, particularly in the case of EUV photoevaporation. These observational constraints, combined with theoretical photoevaporation models, yield mass-loss rates in the range of $\dot{M} \sim 10^{-9} - 10^{-6} M_{\odot} yr^{-1}$ \citep[][]{henney1998, 1999AJ....118.2350H, 2002ApJ...566..315H, 2023ApJ...954..127B, 2024ApJ...967..103B}.

% In EUV-dominated outflows, the mass-loss rate from the disc is controlled by the EUV-driven ionization rate, while FUV heating contributes to the formation of a thin, slowly flowing neutral gas layer in the PDR. 

% EUV photons ionise the hydrogen, heating the gas up to temperatures of $\sim10^4$~K. This ionised region extends down to the so called ionisation front, which, if the EUV flux is sufficiently strong, sits very close to the disc outer radius. The subsonic wind can therefore be launched at the base of the flow. 
% FUV photons instead only dissociate the gas, which is never warmer than $\sim 1000$~K. 

\revision{External photoevaporation also largely affects mass accretion rates. \cite{2017MNRAS.468.1631R} showed that external photoevaporation can drive discs into a late, outside-in clearing phase in which the remaining disc mass is removed faster than it can be replenished by viscous spreading. In this regime the accretion rate does not decline in step with the disc mass, so the accretion rate becomes large compared to the residual disc reservoir, providing a clear signature that external photoevaporation is impacting disc evolution.}

The idea of comparing mass accretion rates and mass-loss rates of protoplanetary discs was already first suggested by \cite{2020MNRAS.497L..40W}. In their work, they provided a framework for testing viscous disc theory by considering stellar accretion and external photoevaporation. Their work highlighted the crucial role of external photoevaporation, particularly in regions like $\sigma$~Orionis, in disc dispersal. By comparing the mass accretion rate onto the star ($\dot{M}_{\rm acc}$) with the mass-loss rate due to photoevaporation ($\dot{M}_{\rm pe}$), they aimed to probe angular momentum transport in the outer disc, a key aspect of viscous theory. Their numerical simulations predicted that in older regions, a correlation between $\dot{M}_{\rm acc}$ and $\dot{M}_{\rm pe}$ would support viscous diffusion, while a significant disparity would suggest otherwise. 

In this work, we refer to the $\sim$1-3 Myr old ONC region, where the majority of the proplyds are, as a case study to explore how externally photoevaporated discs can inform our understanding of angular momentum transport. 
In this work, we explore how externally photoevaporated discs can inform our understanding of angular momentum transport. 
We thus reconsider the direct comparison between mass accretion rates and mass-loss rates from externally induced photoevaporative outflows, for both viscously and MHD-wind-driven evolution. We give a brief overview of disc evolution and the different theories of angular momentum transport in Section~\ref{sec:overview}, with a particular focus on the timescales of the problem in Section~\ref{sec:timescales}. We also extend on the work of \cite{2020MNRAS.497L..40W} by including an analytical derivation for a viscously evolving disc in Section~\ref{sec:theory_pe}, and present the numerical model in Section~\ref{sec:models}. 

\revision{\section{Analytical modelling}}
\label{sec:overview}
\revision{\subsection{Mechanisms of Disc Evolution}}

In the classical picture, angular momentum is redistributed within the disc and, as a consequence, the disc expands. This has a clear observational prediction: the disc outer radius increases over time \revision{\citep[e.g.][]{2018ApJ...864..168N}}. 
Additionally, any standard model of a viscously evolving disc will gradually slow down with time as material spreads to the outer regions, where viscous transport occurs more slowly. Consequently, the ratio of disc mass to mass accretion rate rises as the disc ages \revision{\citep[e.g.][]{Jones2012, 2017MNRAS.468.1631R}}. To ensure that the disc eventually disperses, outflows must be invoked to remove material, as viscosity alone would not result in the disc dispersal \revision{\citep[e.g.][]{Clarke2001, 2014prpl.conf..475A}}. 
% Any viscously evolving disc will also slow down with time - there is more and more material in the outer part of the disc due to spreading, where the evolution is slower. The ratio of disc mass and mass accretion rate increases as the disc gets older. For this reason, we need to couple this evolution with disc winds that can remove material from the disc - otherwise the disc will never disappear.  

% Due to internal winds, a cavity opens and material is removed until an outer disc is left. 

Alternatively to the viscous scenario, MHD-winds are launched along the magnetic field lines threading the disc, removing angular \revision{\citep{2023ASPC..534..465L}}. As a consequence, the disc does not necessarily expand, and the disc outer radius can either remain constant in time or shrink. Additionally, because MHD winds are capable of removing mass from the disc, the mass and the mass accretion rate decrease with time. 
Also a slow down in evolution does not occur in this case, again because the disc does not spread outwards. Hence, the disc evolutionary timescale typically remains constant over time, or can even speed up \citep[e.g.][]{2022MNRAS.512L..74T}. 

% \subsection{External photoevaporation}
% What happens to the disc under the effect of external photoevaporation? 
% Wind driven disc model has no knowledge of what the outer radius of the disc is. it only cares what the local surface density is to set the mass accretion rate. 
% In the case of photoevaporation dominated, high wind rates, photoevaporation is eating into the wind driven discs. but the wind driven disc really doesn't care that I have lost some of the outer material, it doesn't have to communicate with it. the surface density in the inner regions is only declining very very slowly.
How does external photoevaporation impact disc evolution? In an MHD wind-driven disc model, the radius of the outer disc is not a determining factor of its accretion driven evolution, and the mass accretion rate depends solely on the local disc properties (e.g. surface density, magnetic field strength, etc.). When photoevaporation rates are high and dominate the evolution, photoevaporation removes material from the outer disc. However, the MHD wind-driven disc itself remains unaffected by this outer loss, as it operates independently of the outer regions. As a result, the surface density in the inner disc decreases only gradually due to mass-accretion and, hence, the mass accretion rate is not expected to vary significantly over time. 

Conversely, in the viscous picture, external photoevaporative outflows have a much greater impact. 
Rather than expanding, the disc is eroded at its outer edge by the external photoevaporative outflow \revision{\citep{2017MNRAS.468.1631R}}. Since angular momentum transport connects the inner and outer regions, the disc surface density and mass accretion rate both decrease significantly. Due to the nature of viscosity, as the inner material is accreted onto the central star, the outer disc is depleted, leading to a causal connection between mass accretion rates and mass-loss rates within the disc. 

So far, we have neglected the impact of internal photoevaporation, which is driven by radiation from the host star. This assumption is justified under the premise that the mass-loss rates due to externally driven photoevaporative outflows considered in this work are predominantly higher than those resulting from internal photoevaporation \citep{2014prpl.conf..475A, 2017RSOS....470114E}. Additionally, once photoevaporation begins to affect disc evolution, its remaining lifespan is significantly shortened. In environments where external photoevaporation is weak or negligible, internal photoevaporation can play a dominant role in the evolution of the disc.

% The disc instead of expanding is eaten away by the wind, at the outer edge, until it reaches an equilibrium and the disc outer radius almost does not change. Because there is a connection between inner and outer regions due to angular momentum transport, the disc surface density decreases significantly, along with the mass accretion rate. 
% viscous spreading of the disc is impacted by external winds. The disc is truncated, 

\subsection{Timescales} \label{sec:timescales}
Let us now examine the problem in terms of the relevant timescales. 
% The typical viscous timescale for these systems is defined as $t_{\nu} = R^2 / 3\nu(R)$. 
Viscosity is usually assumed to vary with radius in the general form of $\nu \sim R^{\gamma}$. The typical viscous timescale at the outer radius $R_{\rm d}$ is then given by 
\begin{equation}
    t_{\nu} = \frac{R_{\rm d}^2}{\nu(R_{\rm d})} = \frac{R_{\rm d}^2}{\nu_0 R_{\rm d}^{\gamma}}, 
    \label{eq:visc_timescale}
\end{equation}
where $\nu_0$ is constant. 
The viscous timescale, therefore, scales with the disc's size as $t_{\nu} \propto R_{\rm d}^{2-\gamma}$. 
We now want to compare the viscous timescale with the timescale over which the outer radius of the disc evolves due to external photoevaporation, $t_{\rm edge}$. We \revision{begin} by considering the rate of change of the disc mass due to photoevaporation, such that:
\begin{equation}
    \dot{M}_{\rm pe} = \frac{dM_{\rm d}}{dt} = \frac{4 \pi R_{\rm d} \Sigma(R_{\rm d})}{(2-\gamma)} \frac{dR_{\rm d}}{dt},
\end{equation}
where $M_{\rm d}$ is the total disc mass and $\Sigma$ is the disc's surface density, taken to follow the steady-state solution $ \Sigma(R) = \Sigma_{\rm c} (R/ R_{\rm c})^{-\gamma}$, with $\rm \Sigma_c$ the normalisation constant set by the total disc mass and $R_{\rm c}$ is a radial scale factor. Rearranging the equation yields:
\begin{equation}
    \frac{dR_{\rm d}}{dt} = \frac{\dot{M}_{\rm pe}(2-\gamma)}{\revision{4} \pi R_{\rm d} \rm \Sigma_c} \left( \frac{R_{\rm d}}{{\rm R_c}} \right)^{\gamma}.
\end{equation}
This is the rate at which the disc outer edge is evolving under the influence of external photoevaporation, and the corresponding timescale can be defined as
\begin{equation}
    t_{\rm edge} = \frac{R_{\rm d}}{dR_{\rm d}/dt}.
    \label{eq:deltar_timescale}
\end{equation}
The photoevaporative mass-loss rate can be expressed as a function of disc outer radius, $\dot{M}_{\rm pe} \propto R_{\rm d}^{\alpha}$, where $\alpha \sim 1$ for FUV-dominated outflows and $\alpha \sim 3/2$ for EUV-dominated outflows \citep{1998ApJ...499..758J}, provided $R_d>R_g$\footnote{In the case $R_d<R_g$ the mass-loss rate scales exponentially with disc radius, meaning the following argument holds for all $\gamma$.}. This yields a scaling for $t_{\rm edge}$ as a function of disc radius of:  
\begin{equation}
    t_{\rm edge} \propto R_{\rm d}^{2-\gamma-\alpha}. 
\end{equation}
By assuming that the disc viscosity scales linearly with radius ($\gamma =1$, e.g. a constant $\alpha$ disc with $T\propto R^{-1/2}$, \citealt{hartmann1998}), we obtain:
\begin{align}
    & t_{\nu} \propto R_{\rm d}, \\
    & t_{\rm edge, EUV} \propto R_{\rm d}^{-1/2}.
\end{align}
\begin{figure}
    \centering
    \includegraphics[width=\columnwidth,trim=0cm 0cm 0cm 0cm,clip]{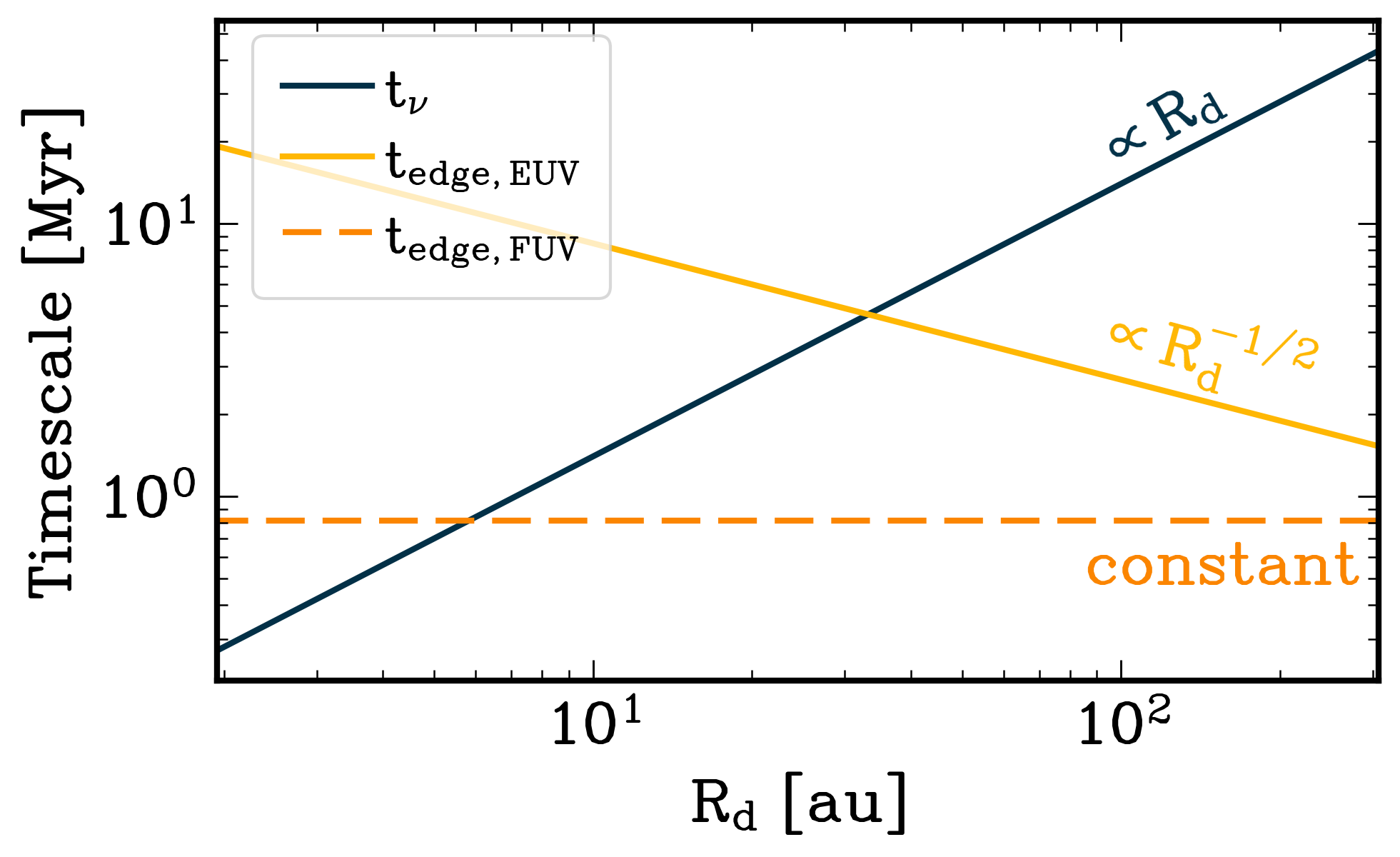}
    \caption{Relevant timescales as a function of disc size for a viscously evolving disc. The viscous timescale (dark navy line) scales as $t_{\nu} \propto R_{\rm d}$, while the timescale for the evolution of the disc's outer radius due to EUV-driven external photoevaporation (yellow solid line) scales as $t_{\rm edge, EUV} \propto R_{\rm d}^{-1/2}$. For an FUV-dominated photoevaporative outflow (orange dashed line), instead, $t_{\rm edge, FUV}$ is independent of $R_{\rm d}$.}
    \label{fig:timescales}
\end{figure}

\noindent We find, instead, that $t_{\rm edge, FUV}$ is a constant function of the disc size. Figure~\ref{fig:timescales} shows these timescales as a function of radius, with the viscous timescale in dark navy and the timescale for the disc outer radius evolution, distinguishing between an EUV and an FUV-dominated outflows, represented by the yellow solid and orange dashed lines, respectively. 
At large radii, $t_{\nu} > t_{\rm edge, EUV}$, which means that external photoevaporation removes material faster than viscous transport can replenish it.  As the size of the disc shrinks due to mass loss, $R_{\rm d}$ decreases and $t_{\rm edge, EUV}$ increases over time. This implies that the rate at which the disc shrinks slows. Simultaneously, $t_{\nu}$ decreases, implying that viscosity becomes relatively more important. Once $R_{\rm d}$ becomes small enough, $t_{\nu} < t_{\rm edge, EUV}$ and the viscous evolution is now faster than the shrinking process. At this point, the disc's is able to reach a pseudo steady-state where its structure is set by viscous redistribution between in the inner and outer edges. At some critical radius, the two timescales are comparable, marking the transition where viscous evolution and mass-loss due to external photoevaporation act on similar timescales. This point also marks a transition from an external photoevaporative outflow dominated contraction to a viscous-dominated evolution as the disc radius shrinks over time. These considerations also hold in the case of mass removal by FUV photoevaporation, although $t_{\rm edge, FUV}$ is independent of $R_{\rm d}$ at large radius. 
More generally, the transition from a photoevaporation dominated regime to a viscous-dominated regime occurs when $2-\gamma-\alpha \leq 0$ and $2-\gamma > 0$. These conditions allow for an arbitrary choice of viscosity prescription, provided that $2-\alpha \leq \gamma < 2$ (a condition most standard viscous disc models satisfy), although this requirement disappears inside $R_g$ where the viscous timescale will always eventually become shorter than the disc radius evolution timescale.

\subsection{Viscous discs evolve towards an accretion rate that matches the external photoevaporation rate} \label{sec:theory_pe}

The evolution of the timescales indicates that sufficiently strong external photoevaporation will always drive the disc towards a state where the viscous timescale at the outer edge is shorter than the timescale on which the outer edge is evolving due to photoevaporation. In this case, we can consider the disc to be evolving as if the outer edge remains fixed.
With this general understanding in mind, we now approach the problem from an analytical perspective. Following \cite{2004ApJ...605..880R}, we consider the specific case of a viscous disc where viscosity scales linearly with radius, $\nu = \nu_0 R$ \footnote{We note our argument completely generalises to other viscosity laws, we choose $\nu\propto R$ simply for algebraic simplicity.}. The dimensionless evolution equation is
\begin{equation}
    \frac{\partial S}{\partial \tau} - \frac{3}{4} \frac{\partial^2 S}{\partial x^2} = 0.
    \label{eq:dimentionless_evolution}
\end{equation}
This is a standard diffusion equation, where the dimensionless spatial and temporal variables are $x = (R/R_{\rm g})^{1/2}$ and $\tau = t/t_{\rm vis}$, with the viscous timescale defined at $R_{\rm g}$ as $t_{\rm vis} = R_{\rm g}^2 / \nu_g$. $S$ is the angular momentum flux and it depends on the disc's surface density, $S = \Sigma x^3$. 
We can find the general solution via separation of variables, by assuming that $S(x,\tau) = X(x) T(\tau)$. This yields:
\begin{align}
    & T(\tau) = T_0 e^{- \lambda \tau}, \\
    & X(x) = A {\rm sin}(kx) + B {\rm cos}(kx), \quad k^2 = \frac{4}{3}\lambda.
\end{align}
By adopting the assumption of zero torque at the origin, as in \cite{1974MNRAS.168..603L}, we impose $S=0$ at $R=0$, and find the trivial solution $B=0$, since $k \neq 0$. The solution takes the form
\begin{equation}
    X(x) = A {\rm sin}(kx).
\end{equation}
We now include the effect of external photoevaporation. Which in the limit where the disc outer radius evolves slower than the viscous timescale, allows us to approximate the disc outer radius as fixed.  This approximation allows us to simplify the model while correctly capturing the large-scale evolution of the system. Additionally, while this applies in the case of external photoevaporation, it would not apply for internally driven photoevaporative outflows, as the removal timescale is different and an additional term must be included in the evolution equation (Eq.~\ref{eq:dimentionless_evolution}).
Thus, if we consider that external photoevaporative outflows are acting on the disc at the outer radius, we can safely assume that the disc surface density drops quickly at $R = R_{\rm d}$, and set $S=0$. This implies $A {\rm sin}(kx_{\rm d}) = 0$, which means $kx_{\rm d} = n\pi$ for $n=1,2..$.
Thus, we expand $S(x,\tau)$ as the sum of sine modes, and the general solution becomes
\begin{equation}
    S(x,\tau) = \sum_{n=1}^{\infty} A_n {\rm sin}(k_n x) e^{-\lambda_n \tau},
\end{equation}
where $k_n = n \pi / x_{\rm d}$ and $\lambda_n = 3/4 k_n^2$. The method of calculation of the coefficients $A_n$ are shown in Appendix~\ref{appx:coefficients_an}. Figure~\ref{fig:analytical_solution} shows the analytical solution to the problem at different time steps, illustrating the effects of an external wind at the outer edge of the disc. 
\begin{figure}
    \centering
    \includegraphics[width=\columnwidth,trim=0cm 0cm 0cm 0cm,clip]{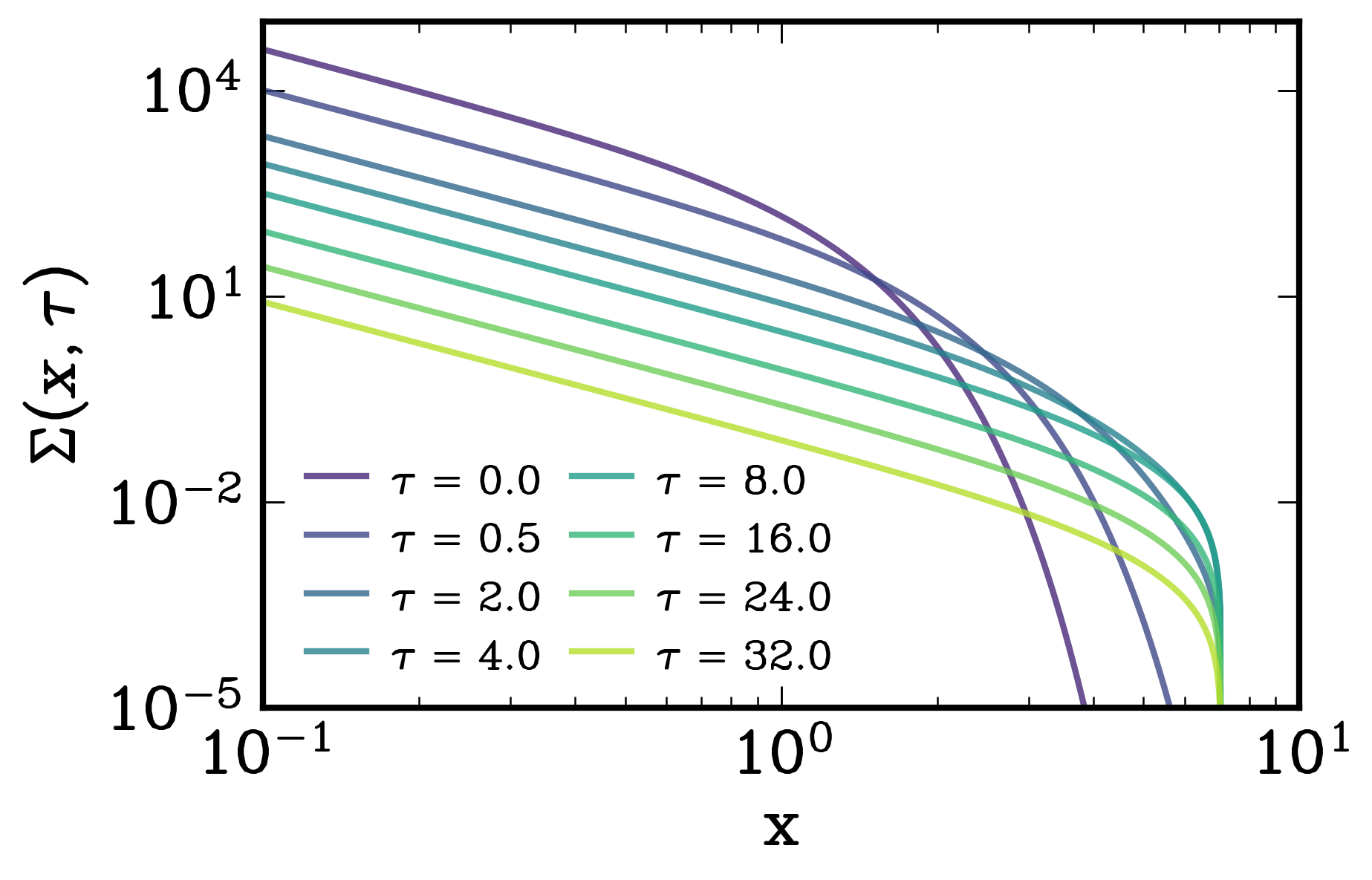}
    \caption{Disc surface density as a function of $x$, at different time steps. It comes from the analytic solution to Eq.~\ref{eq:dimentionless_evolution}, illustrating the effect of an external photoevaporative outflow at the outer boundary.}
    \label{fig:analytical_solution}
\end{figure}

The mass flux through the disc is defined as \citep[see Eq.~9 in][]{2004ApJ...605..880R}:
\begin{equation}
    \dot{M} = 6 \pi R^{1/2} \frac{\partial }{\partial R}\left(\Sigma_g \nu_g R^{1/2}\right).
\end{equation}
Since our disc will have reached a pseudo-steady state after a few local viscous timescales at the outer edge, the higher order terms in the series rapidly decay away becoming negligible, and we are left with:

\begin{equation}
    S(x,\tau)\approx A_1\sin(\pi x) \exp\left(-\pi^2\tau\frac{3R_g}{4R_d}\right)
\end{equation}

Therefore, evaluating the mass flux at the origin, we find the accretion rate onto the star as:
\begin{equation}
    \dot{M}_{\rm acc} = -\dot{M}(x=0) \approx 3 \pi^2 \nu_0 A_1 \frac{R_g^{3/2}}{R_{d}^{1/2}} \exp\left(-\pi^2\tau\frac{3R_g}{4R_d}\right)
\end{equation}

Now, evaluating at the outer edge, where by mass continuity we know the mass-flux in the wind must equal the mass-flux through the disc at $R=R_{d}$, we find:
\begin{equation}
        \dot{M}(R=R_d) \approx - 3 \pi^2 \nu_0 A_1 \frac{R_g^{3/2}}{R_{d}^{1/2}} \exp\left(-\pi^2\tau\frac{3R_g}{4R_d}\right) = - \dot{M}_{\rm acc} = - \dot{M}_{\rm pe}
\end{equation}
%By scaling times similarly to the mass accretion rate, we can now write the mass-loss rate as
%\begin{equation}
%    \begin{split}
%        \dot{M}_{\rm pe} & = -2 \pi \frac{R_{\rm g}^2}{t_{\rm vis}}\int_0^{x_{\rm d}} x^3 \frac{\partial S}{\partial \tau} dx \\
%        & = -4 \pi \nu_{\rm g} \sum_n A_n (-\lambda_n) \frac{{\rm sin}(k_n x_{\rm d})}{k_n} e^{-\lambda_n \tau} \\
%        & = 3 \pi \nu_{\rm g} \sum_n A_n k_n (-1)^n e^{-\lambda_n \tau} = \dot{M}_{\rm acc}.
%    \end{split}
%\end{equation}
An important consequence of the above approximation is that, in order to maintain the disc's outer radius constant over time, all the mass flux reaching the edge is lost to the photoevaporative outflow. This implies that the mass-loss rate is equal to the mass accretion rate $\dot{M}_{\rm pe} = \dot{M}_{\rm acc}$. 
% This conclusion emphasizes that the mass accretion rate is expected to scale proportionally to the mass-loss rate. 
% Furthermore, since the mass-loss rate depends on the disc size, the same scaling naturally extends to the mass accretion rate. 
This conclusion emphasizes the role of viscosity in accretion discs. Viscosity, in fact, is responsible for redistributing angular momentum within the disc, therefore connecting the innermost regions with the outer part of the disc. To accrete, an externally photoevaporating disc must supply the outflow with the angular momentum flux necessary to accrete onto the star, and thus must match the accretion rate and outflow rate on timescales longer than the viscous time.  However, this result holds only for a disc evolving under viscosity, whereas the case of an MHD wind-driven disc is more complex. The nature of an MHD wind means the different radii are not all casually coupled (the MHD wind term is hyperbolic, rather than parabolic as in the viscous case). There is no direct casual connection between the outer disc and inner disc, and as such the accretion rate onto the star is set by the local conditions in the inner disc, which are not directly affected by the outer disc's radius.  
We also note that we have made the assumption that viscosity scales linearly with radius to make the algebra easier, and different viscosity laws should yield a similar result due to angular momentum conservation: the accretion rate and mass-loss rate in an external photoevaporative wind should be tightly correlated once the evolution timescale for the outer radius is longer than the local viscous timescale. A problem where one would allow the outer disc edge to also evolve is closely related to ``Stefan Problems'' \citep[e.g.][]{Gupta2012}. Analytical investigations that turn the case of an externally photoevaporating disc into such a Stefan problem may yield further insight. In particular, in the case of a fixed outer edge, the accretion rate always matches the photoevaporation rate at late times. However, our numerical calculations show that due to the evolving outer edge the accretion rate is always slightly lower than the photoevaporation rate at late times, as the disc always contracts at late times.

\section{Numerical Modelling} \label{sec:models}
We now numerically solve the equation of mass and momentum conservation for a disc. Thus, we model a protoplanetary disc undergoing viscous evolution by employing the 1D code developed in \cite{2017MNRAS.469.3994B}. The MHD wind driven evolution is implemented following \cite{2022MNRAS.514.1088Z} and the original prescription is described in \cite{2022MNRAS.512.2290T}. The disc evolves following the general equation:
\begin{equation}
    \begin{split}
        \dfrac{\partial\Sigma}{\partial t} = & \dfrac{3}{R} \dfrac{\partial}{\partial R} \left[\frac{1}{r\Omega} \dfrac{\partial}{\partial R} \left( r^2 \alpha_{\rm SS} \Sigma c_{\rm s}^2 \right)\right] - \dot{\Sigma}_{\rm pe} \\ 
        & + \dfrac{3}{2R} \dfrac{\partial}{\partial R} \left[ \dfrac{\alpha_{\rm DW} \Sigma c_{\rm s}^2}{\Omega}\right]
        - \dfrac{3 \alpha_{\rm DW} \Sigma c_{\rm s}^2}{4(\lambda-1) R^2 \Omega},
    \end{split}
    \label{eq:master_eq}
\end{equation}
where $\Omega$ is the Keplerian angular frequency. We assume a locally isothermal gas, with sound speed $c_{\rm s}(R)$. The first term in the equation describes the evolution of the disc governed by viscous processes regulated by turbulence. 
According to the \cite{1973A&A....24..337S} $\alpha$-prescription, the kinematic viscosity has the form $\nu = \alpha_{\rm SS} c_{\rm s} H$ \citep[e.g.][]{1974MNRAS.168..603L}, where $ \alpha_{\rm SS}$ is a dimensionless parameter that scales with the square of the turbulent velocity (in units of the sound speed) and $H$ is the disc's scale height. 
Under these conditions, the disc evolves following the analytical self-similar solutions of \cite{1974MNRAS.168..603L}, if the viscosity law is self-similar. Part of the disc material is accreted onto the central star, while the angular momentum within the disc is redistributed, and the net result is a viscous spreading of the whole disc. 

The second term in the equation is the rate at which mass is removed from the outer disc radius due to external photoevaporative outflows (see the next Section). 

The third and last terms in Eq.~\ref{eq:master_eq} include the contribution of the MHD wind. The parameter $\alpha_{\rm DW}$ quantifies the angular momentum extracted by the MHD wind, and therefore regulates the advection of gas. The parameter $\lambda$, represents the magnetic lever arm, which ultimately determines the MHD wind's mass-loss rate due to the magnetic field and whose precise value is still unconstrained.  
%which ultimately determines the wind mass-loss rate. 
In the formulation presented by \cite{2022MNRAS.512.2290T}, $\alpha_{\rm DW}$ is defined similarly to the original parametrization of $ \alpha_{\rm SS}$. Just as the viscous accretion rate scales with $\alpha_{\rm SS}$, so the MHD wind-driven accretion rate is proportional to $\alpha_{\rm DW}$, allowing a straightforward comparison between these two coefficients. Hence, \cite{2022MNRAS.512.2290T} introduce the parameter $\psi$, which quantifies the MHD wind's intensity,
\begin{equation}
    \psi \equiv \frac{\alpha_{\rm DW}}{\alpha_{\rm SS}}.
\end{equation}
For a disc that evolves purely viscously, $\psi=0$, while MHD winds dominate the evolution of the disc for $\psi \rightarrow \infty$. In this work, we consider two sets of solutions: the purely viscous evolution, with a constant $\alpha_{\rm SS}$ and $\psi=0$ (hence $\alpha_{\rm DW}=0$), and the MHD wind-driven evolution, with constant $\alpha_{\rm DW}=10^{-3}$ and $\psi=10^4$ \citep{2022MNRAS.512.2290T, 2022MNRAS.514.1088Z}. We also consider the specific case of $\lambda=3$, as a representative intermediate value within the claimed observed range of $\lambda$ from 1.6 to 5 \citep{2022MNRAS.512.2290T, 2022MNRAS.514.1088Z}. 

\subsection{Including external photoevaporation} \label{sec:model_extpe}
The disc is subject to intense UV stellar irradiation, which is typically coming from massive O or B-type stars. As mentioned in Section~\ref{sec:intro}, both EUV and FUV photons contribute to driving a photoevaporative outflow, and the results illustrated in Section~\ref{sec:theory_pe} are valid for both FUV- and EUV-dominated cases. However, this work focuses solely on EUV photoevaporation. This choice is motivated by the fact that EUV-driven mass-loss rates are easier to measure through radio emission than the FUV case, leading to more direct and robust measurements. EUV photons have sufficient energy to ionize hydrogen-rich gas, which then produces free-free emission. This emission is directly related to the density of the outflow \citep[e.g.][]{Churchwell1987}, making it easier to measure using radio observations. This is particularity true for resolved multi-wavelength observations where the disc and outflow can be cleanly separated and optical depth effects can be taken into account \citep{2023ApJ...954..127B}.    

We therefore include EUV external photoevaporation in our models, by removing material from the disc outer edge. We follow the analytic formulation of \cite{1998ApJ...499..758J} for EUV-dominated outflows, where the mass-loss rate is given by:
\begin{equation}
    \frac{\dot{M}_{\rm pe}}{M_{\odot} {\rm yr}^{-1}} = 1.78 \times 10^{-7} \left( \frac{d}{0.1 \, \rm pc} \right)^{-1} \left( \frac{\Phi}{10^{49} \, \rm s^{-1}} \right)^{1/2} \left( \frac{R_{\rm d}}{100 \, \rm au} \right)^{3/2} .
    \label{eq:mdot_euv}
\end{equation}
Here, $d$ is the distance from the external massive star, $\Phi$ is the ionizing flux and $\rm R_d$ is the disc's outer radius. 
% $f$ is the fraction of ionizing photons that penetrates the flow.
Notably, the EUV driven mass-loss rate scales super-linearly with the disc outer radius. 
External photoevaporation is implemented in our numerical model following the approach of \cite{2007MNRAS.376.1350C}.
% We emphasize that, as the disc outer radius evolves over time, the EUV photoevaporation mass-loss rate is re-calculated at each step and varies accordingly. 

\section{Results} \label{sec:results}
We initialise all our calculations (viscous or MHD wind) with a zero-time similarity solution from \cite{1974MNRAS.168..603L} which has the following form
\begin{equation}
    \Sigma (R, t=0) = \frac{M_0}{2 \pi R_0 R} {\rm exp} \left( - \frac{R}{R_0} \right)
\end{equation}

\subsection{Evolution of disc properties}
\label{sec:disc_properties}
To gain insight into the theoretical framework, we initially examine the evolution of the properties of a sample disc. We evolve a disc with an initial mass of $M_0=0.2$~M$_{\odot}$ and an initial scaling radius of $R_0=20$~au, hosting a solar-mass star at the centre. We choose a temperature distribution where the sound speed scales with the power-law index $R^{-1/4}$. 

\begin{figure}
    \centering
    \includegraphics[width=\columnwidth,trim=0cm 0cm 0cm 0cm,clip]{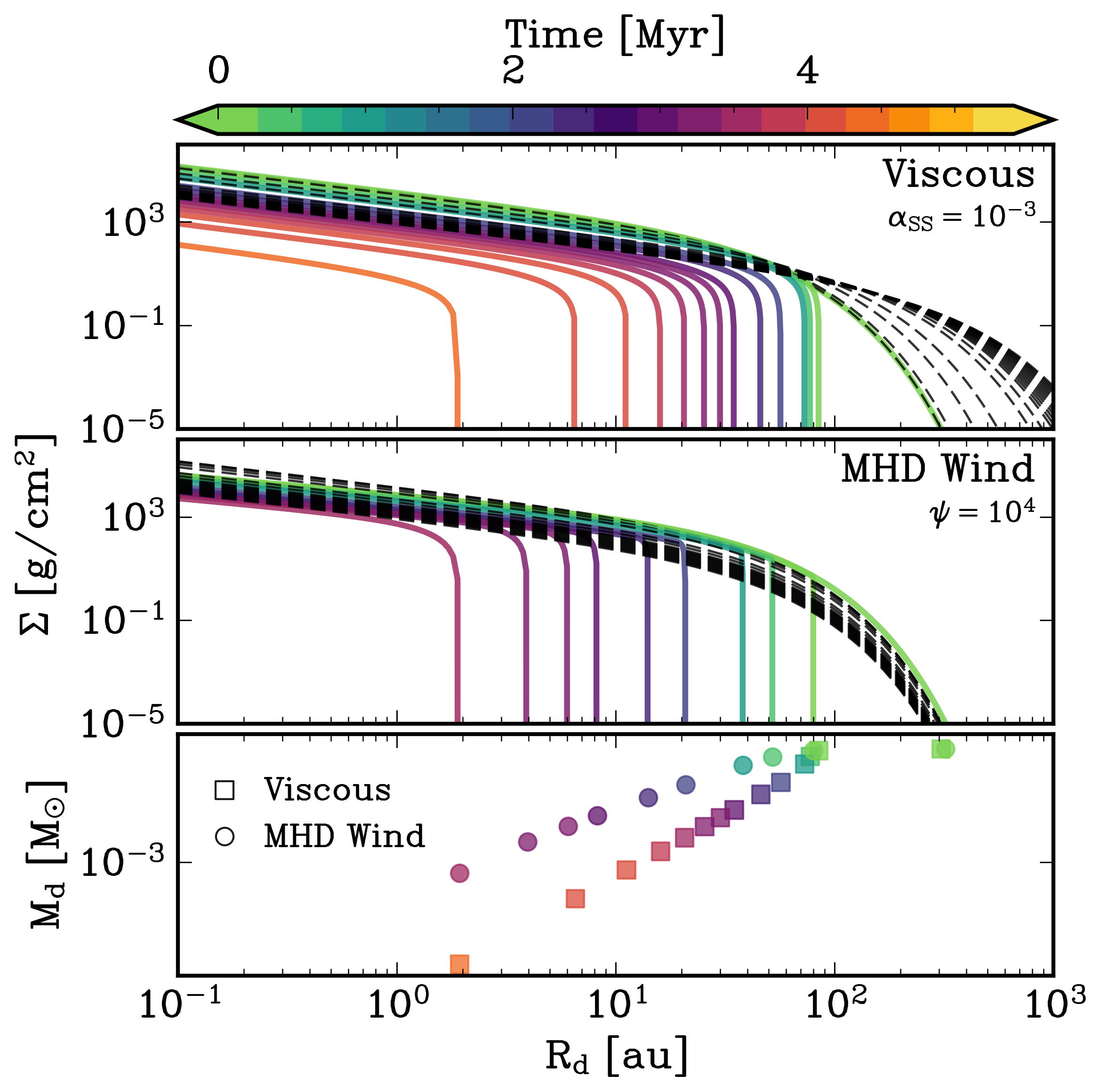}
    \caption{Evolution of the disc surface density profiles for the viscous ($\alpha_{\rm SS}=10^{-3}$) and wind-driven scenarios (top and middle panels, respectively) for an initial disc mass of 0.2~M$_{\odot}$ and scaling radius of 20~au. The disc is located at 0.3~pc from the irradiating source.  
    The dashed black lines in the top and middle panels indicate the evolution of the disc in the absence of external photoevaporation, for each corresponding scenario. 
    The bottom panel shows the evolution of the disc radius and disc mass: squares are used for the viscously evolving discs and circles for the wind-driven discs. Coloured lines and points correspond to $t=0, 0.1, 0.5, 1.0, 2.0, 2.5, 3.0, 3.2, 3.4, 3.6, 3.8, 4.0, 4.2, 4.4$~Myr.}
    \label{fig:disc_evolution_0.3pc_alpha1e-3}
\end{figure}

\begin{figure}
    \centering
    \includegraphics[width=\columnwidth,trim=0cm 0cm 0cm 0cm,clip]{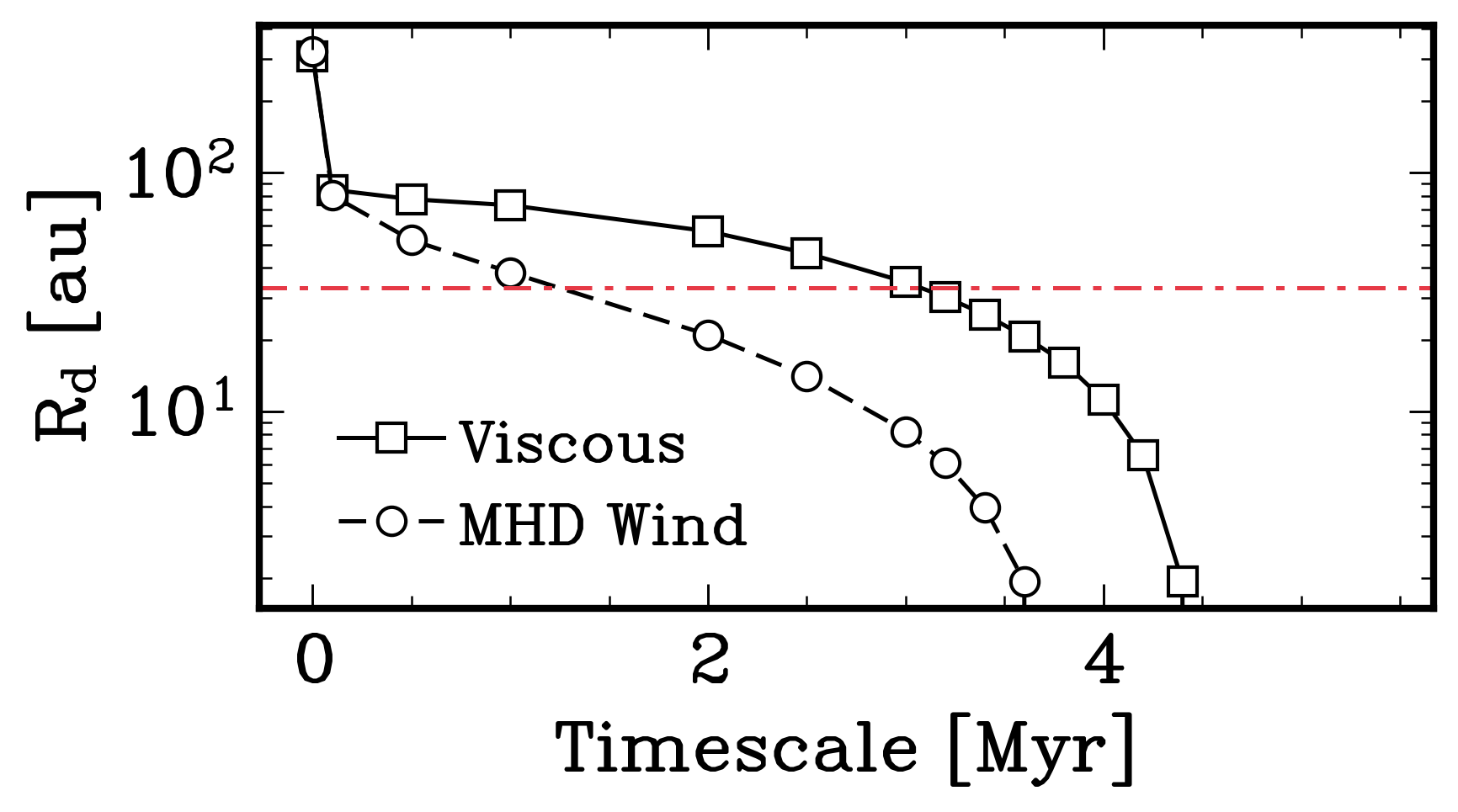}
    \caption{Evolution of the disc outer radius for the viscous and MHD wind-driven discs presented in Figure~\ref{fig:disc_evolution_0.3pc_alpha1e-3}. The red dash-dotted line marks the location at which the viscous disc transitions from a photoevaporation-dominated regime to viscous-dominated evolution (see Figure~\ref{fig:timescales} and Section~\ref{sec:timescales}).}
    \label{fig:evolution_rout}
\end{figure}

An example of the evolution of the disc's surface density for the viscous and MHD wind-driven scenarios is illustrated in the top two panels of Figure~\ref{fig:disc_evolution_0.3pc_alpha1e-3}. Both discs are located at 0.3~pc from the external irradiating source, with an EUV photon luminosity of $\Phi = 10^{49}$~s$^{-1}$. The adopted ionising photon luminosity $\Phi=10^{49}\,\mathrm{s^{-1}}$ is representative of an O-type massive star in the ONC, comparable to $\theta^1$C Ori \citep[spectral type $\sim$O6--O7, ][]{2005ApJS..160..379F}.
In the absence of an external photoevaporative outflow, a viscous disc would spread outward and the outer radius of the disc would increase over time, as highlighted by the dashed black lines in the top panel. Here, in contrast, material within the disc is removed via photoevaporation and the disc shrinks instead. The combined effects of an MHD wind and an external photoevaporative outflow result in a qualitatively similar evolution in this case, with the disc shrinking over time. The evolution of the disc's outer radius at every time step is illustrated in Figure~\ref{fig:evolution_rout}, with the distinction between the viscous and MHD wind cases. The red dash-dotted line marks the transition from a photoevaporation-dominated to a viscous-dominated evolutionary regime (see Figure~\ref{fig:timescales} and Section~\ref{sec:timescales}). Initially, the outer disc radius decreases gradually at an almost constant rate, but once the viscous evolution becomes important, its contraction accelerates significantly, as the viscous time becomes shorter and shorter as the radius shrinks. 
Here, we have selected a disc with typical initial values and the exact evolution of disc mass and disc outer radius in the two scenarios depends on the choice of such initial conditions. Specifically, we find that the local surface density drops by almost three orders of magnitude over $\sim 4.5$~Myr in the viscous scenario, while it decreases by less than an order of magnitude over $\sim 3$~Myr in the MHD wind-driven case. 
% External photoevaporation more significantly depletes a viscously evolving disc, compared to a disc dominated by an MHD wind. The overall effect is very similar to a disc evolving under the effect of both an MHD wind and an external wind. The disc outer radius does not differ significantly in these two cases, as shown in the bottom panel of Figure~\ref{fig:disc_evolution_0.3pc_alpha1e-3}: there is only a minor difference of about $\lesssim 1$~au by the time the two discs reach $\sim 4.5$~Myr, after which the disc is completely depleted of material. 
Nonetheless, we always observe that the evolution of the disc surface density differs quite significantly between a viscous and a MHD wind-driven disc. A viscously evolving disc undergoes a more dramatic decrease in the local surface density over the disc lifetime, particularly in the innermost regions, compared to its MHD wind-driven counterpart. In the latter, in fact, the evolution of the surface density is much slower, with a minor decrease over a smaller timescale. This is due to the intrinsic nature of a viscously evolving disc: the redistribution of angular momentum within the disc causes the disc surface density to lose mass faster as it can transport more material into the photoevaporative outflow, compared to a case in which external photoevaporation is absent, as illustrated by the top panel of Figure~\ref{fig:disc_evolution_0.3pc_alpha1e-3}. The disc needs to replenish the outer regions that are constantly depleted of material.

\begin{figure}
    \centering
    \includegraphics[width=\columnwidth,trim=0cm 0cm 0cm 0cm,clip]{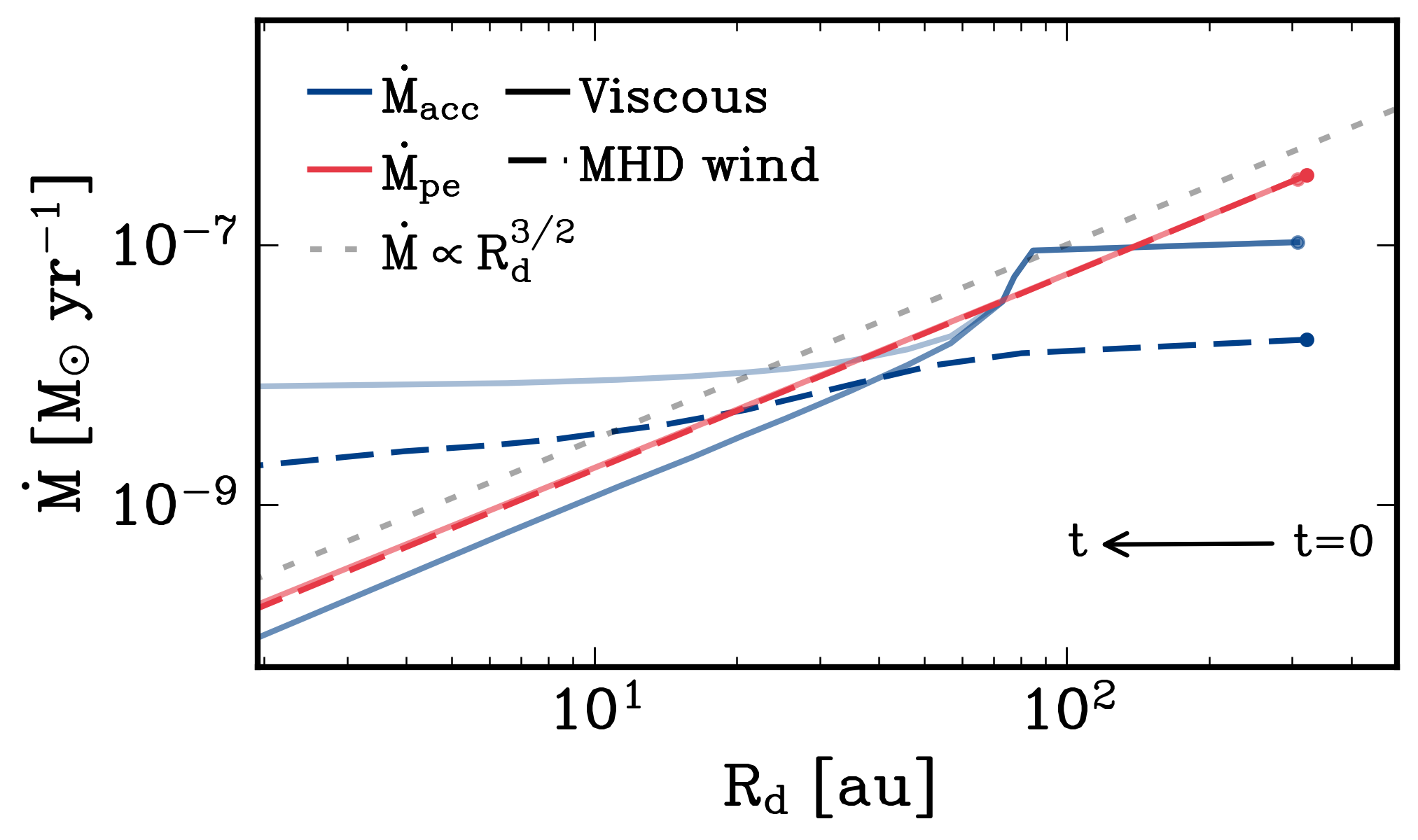}
    \caption{Evolution of the mass accretion (blue) and photoevaporative mass-loss (red) rates with respect to the disc size, for the disc presented in Figure~\ref{fig:disc_evolution_0.3pc_alpha1e-3}. Solid and dashed lines represent the results for the viscous and MHD wind-driven disc, respectively. The blue solid line with lower opacity indicates the values of mass accretion rates for a disc that is not undergoing external photoevaporation.  The grey dotted line shows the scaling $\dot{M} \propto R_{\rm d}^{3/2}$. The black arrow indicates time evolution, with the coloured dots highlighting the $t=0$ point.}
    \label{fig:mdot_rd_0.3pc_alpha1e-3}
\end{figure}

The rapid variation in surface density directly impacts the mass accretion rate of the disc. Figure~\ref{fig:mdot_rd_0.3pc_alpha1e-3} illustrates the relationship between the mass accretion and photoevaporation rate. This plot reflects the theoretical predictions derived in Section~\ref{sec:theory_pe}. Although the photoevaporative mass loss rates of both a viscous and wind-driven discs must follow the $R_{\rm d}^{3/2}$ scaling, marked by the grey dotted line (see Eq.~\ref{eq:mdot_euv}), the behaviour of the mass accretion rates is significantly different. In a viscously evolving disc, the mass accretion rate initially remains almost unchanged (consistent with evolution when the age is less than the disc's initial viscous timescale) but then begins to decreases once the age exceeds the disc's initial viscous timescale, following the same scaling with the radius of the disc ($R_{\rm d}^{3/2}$). Without external photoevaporation, the disc would maintain this steady behaviour and the accretion rate would remain almost constant with disc size, as indicated by the blue solid line with slightly lower opacity. 
In contrast, in an MHD wind-dominated disc, the evolution is entirely independent of the outer regions, and the mass accretion rate slowly evolves as the disc's size evolves, as illustrated by the blue dashed line in Figure~\ref{fig:mdot_rd_0.3pc_alpha1e-3}. However, this correlation is implicit and does not result from a physical link between accretion and disc size. Rather, both naturally decrease with time, independently of each other. 

\subsubsection{The effect of a lower $\alpha$ viscosity} 
\label{sec:lower_viscosity}

\begin{figure}
    \centering
    \includegraphics[width=\columnwidth,trim=0cm 0cm 0cm 0cm,clip]{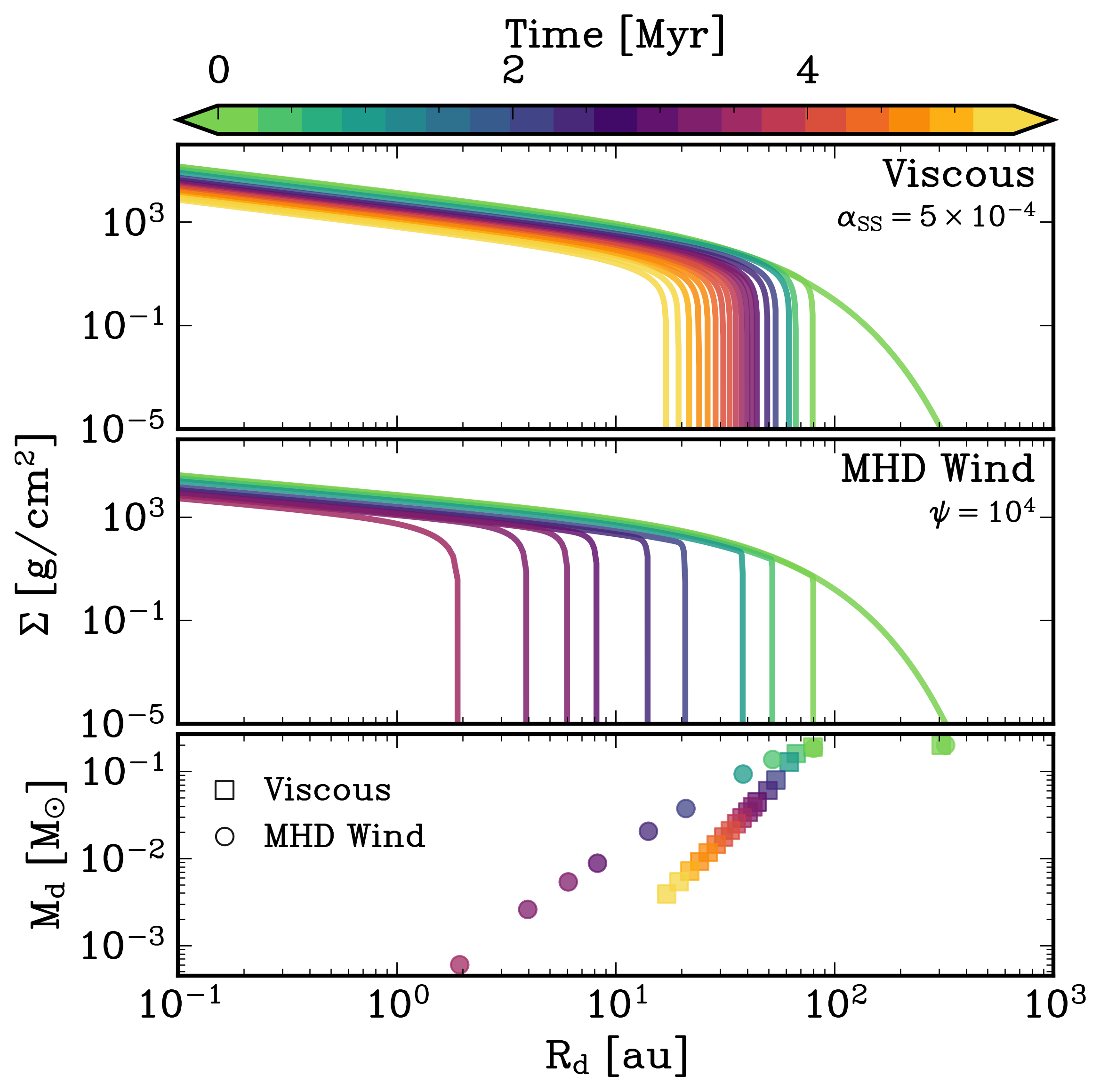}
    \caption{Same as Figure~\ref{fig:disc_evolution_0.3pc_alpha1e-3} but the viscous disc now evolves with a lower viscosity, $\alpha_{\rm SS}=5 \times 10^{-4}$. The MHD wind-driven evolution is the same as previously shown.}
    \label{fig:disc_evolution_0.3pc_alpha5e-4}
\end{figure}

The viscous time scale sets the time over which the disc surface density evolves and, more specifically, the viscous timescale is inversely proportional to the viscosity within the disc. As a consequence, discs with lower viscosity evolve on a longer timescale. The evolution of the two discs is significantly different. This is shown in Figure~\ref{fig:disc_evolution_0.3pc_alpha5e-4}, where the viscous disc has a lower turbulent viscosity, $\alpha_{\rm SS}=5 \times 10^{-4}$, compared to previous results presented in the upper panel of Figure~\ref{fig:disc_evolution_0.3pc_alpha1e-3}. After $\sim 5$~Myr, the viscous disc still has an outer radius $\gtrsim 10$~au and the surface density has decreased by almost two orders of magnitude. The MHD wind-dominated disc is much smaller in comparison and the gas is depleted relatively quickly. 
The main consequence for disc evolution is that a viscous disc will transition from a photoevaporation-dominated regime to a viscous-dominated regime slightly later than it would in a case with higher turbulent viscosity. Thus, the trends shown in Figure~\ref{fig:mdot_rd_0.3pc_alpha1e-3} remain valid, with the turn over in the accretion rate occurs only slightly later. Therefore, our inferences about the differences in the qualitative evolution of a viscous and MHD-wind driven disc do not depend on the assumed strength of the viscosity.

% The disc outer radii start to differ quite significantly in the two scenarios, especially after 1 Myr. Consequently, the total disc mass also remains noticeably different between viscous and wind-driven evolution, at every time step. 

\subsubsection{The effect of a weaker photoevaporation rate} 
\label{sec:larger_distance}

% \gb{disc surface density has decreased only by 1 order of magnitude }

\begin{figure}
    \centering
    \includegraphics[width=\columnwidth,trim=0cm 0cm 0cm 0cm,clip]{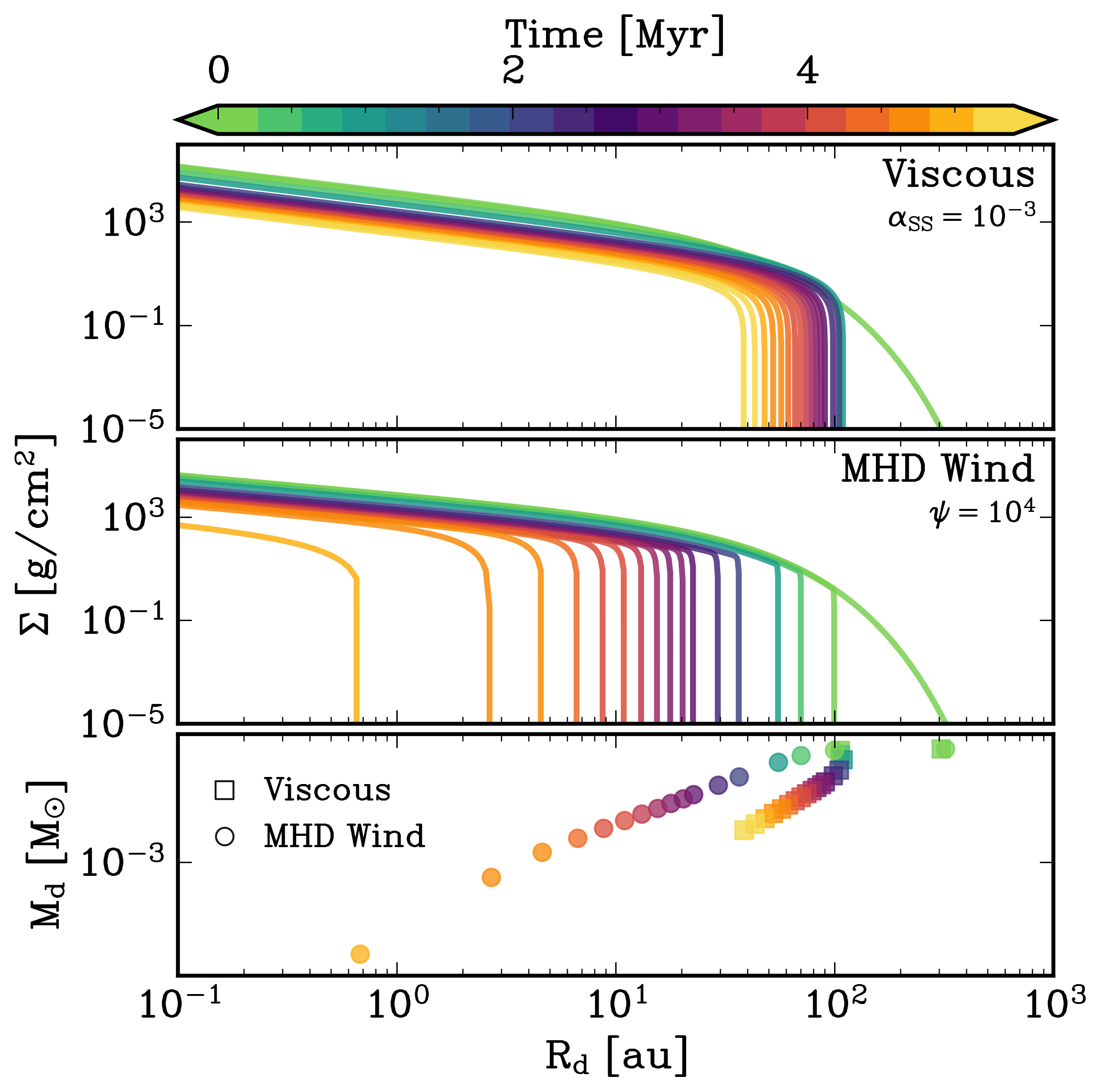}
    \caption{Same as Figure~\ref{fig:disc_evolution_0.3pc_alpha1e-3} but for a disc located at 1~pc from the irradiating source.}
    \label{fig:disc_evolution_1.0pc_alpha1e-3}
\end{figure}

As one might expect, protoplanetary discs that sit farther away from the irradiating source are less impacted by external photoevaporation. Figure~\ref{fig:disc_evolution_1.0pc_alpha1e-3} shows the evolution of the same disc presented in Figure~\ref{fig:disc_evolution_0.3pc_alpha1e-3}, but located at 1.0~pc from the external star. Both discs evolve much slower over similar timescales; hence, they are overall larger and more massive, ultimately surviving until 5~Myr. A weaker photoevaporative outflow has an effect on a viscous disc similar to that of a lower turbulent viscosity (compare the upper panel of Figure~\ref{fig:disc_evolution_0.3pc_alpha5e-4}). The disc outer radius evolves on a much longer timescale, and, as a consequence, the disc has enough time to viscously respond to changes to the outer boundary. Unsurprisingly, when the effect of external photoevaporation is reduced, the evolution is primarily governed by either viscosity or MHD winds. 

\revision{As mentioned in Section~\ref{sec:overview}}, we neglect internal photoevaporation. However, internal photoevaporation will not alter the scenario presented here as long as the mass loss rate is sufficiently high for the disc to transition to a region where $t_{\nu} < t_{\rm edge}$, before internal photoevaporation can clear the disc. 
%Although this regime warrants further investigation, a more detailed analysis falls beyond the scope of this paper and is left for future work. 

\subsection{Evolution of mass accretion rates and photoevaporative mass-loss rates}
\label{sec:macc_mwind}

\begin{figure}
    \centering
    \includegraphics[width=\columnwidth,trim=0cm 2.15cm 0cm 0cm,clip]{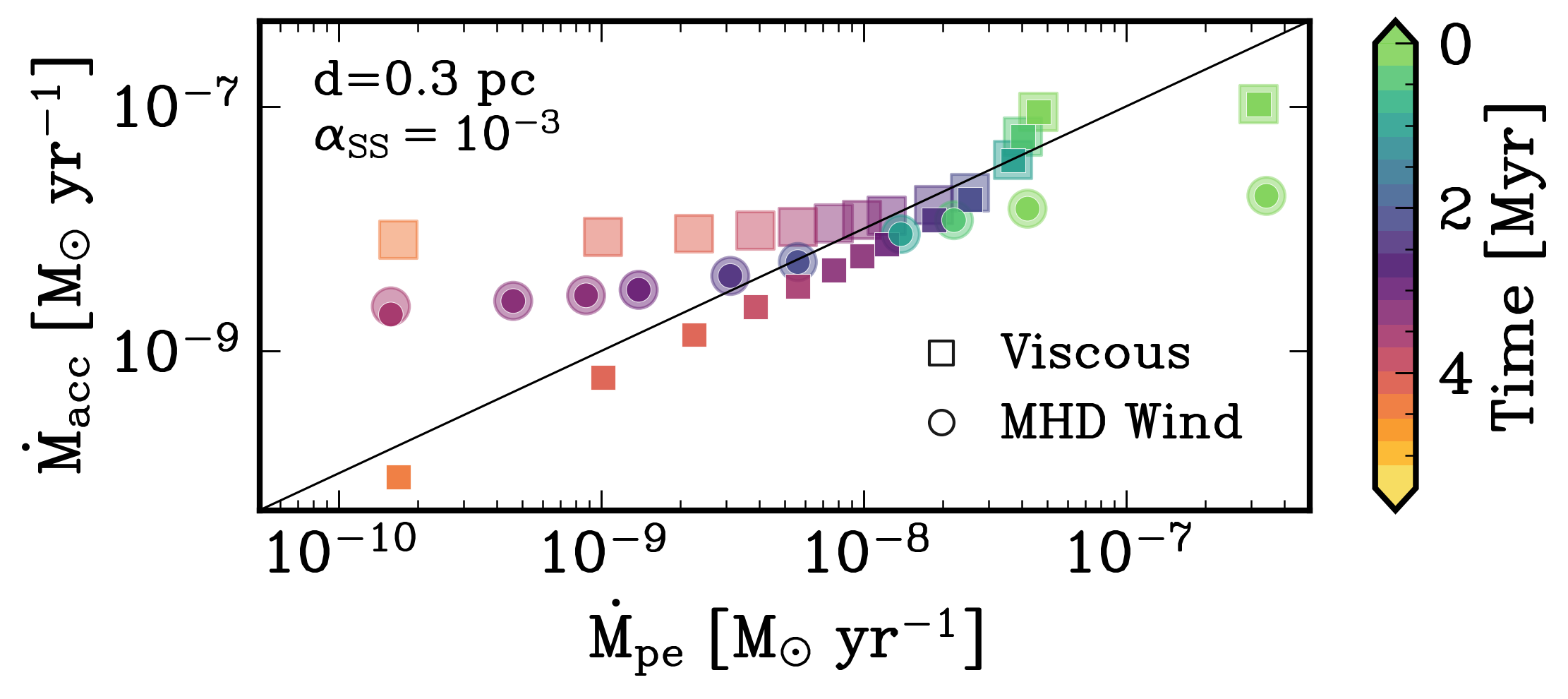}
    \includegraphics[width=\columnwidth,trim=0cm 2.15cm 0cm 0cm,clip]{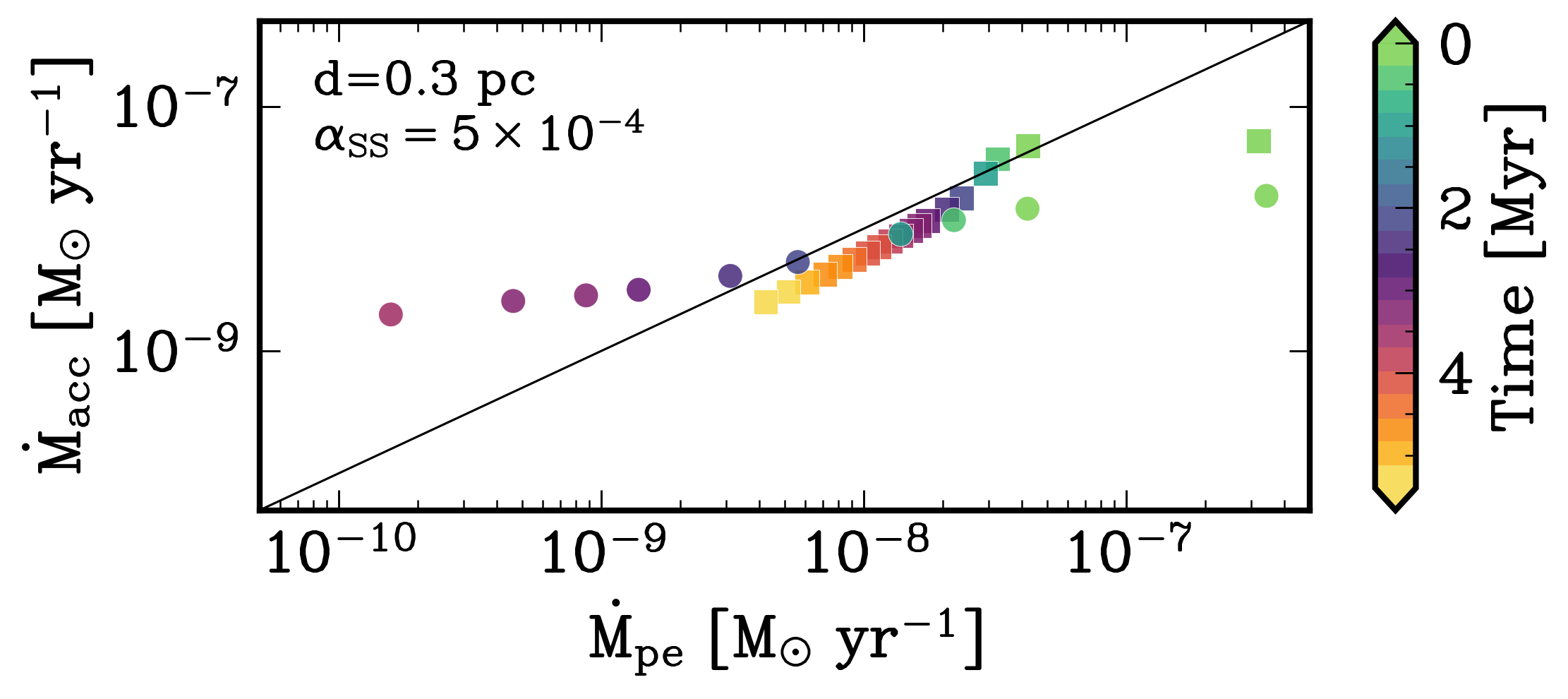}
    \includegraphics[width=\columnwidth,trim=0cm 0cm 0cm 0cm,clip]{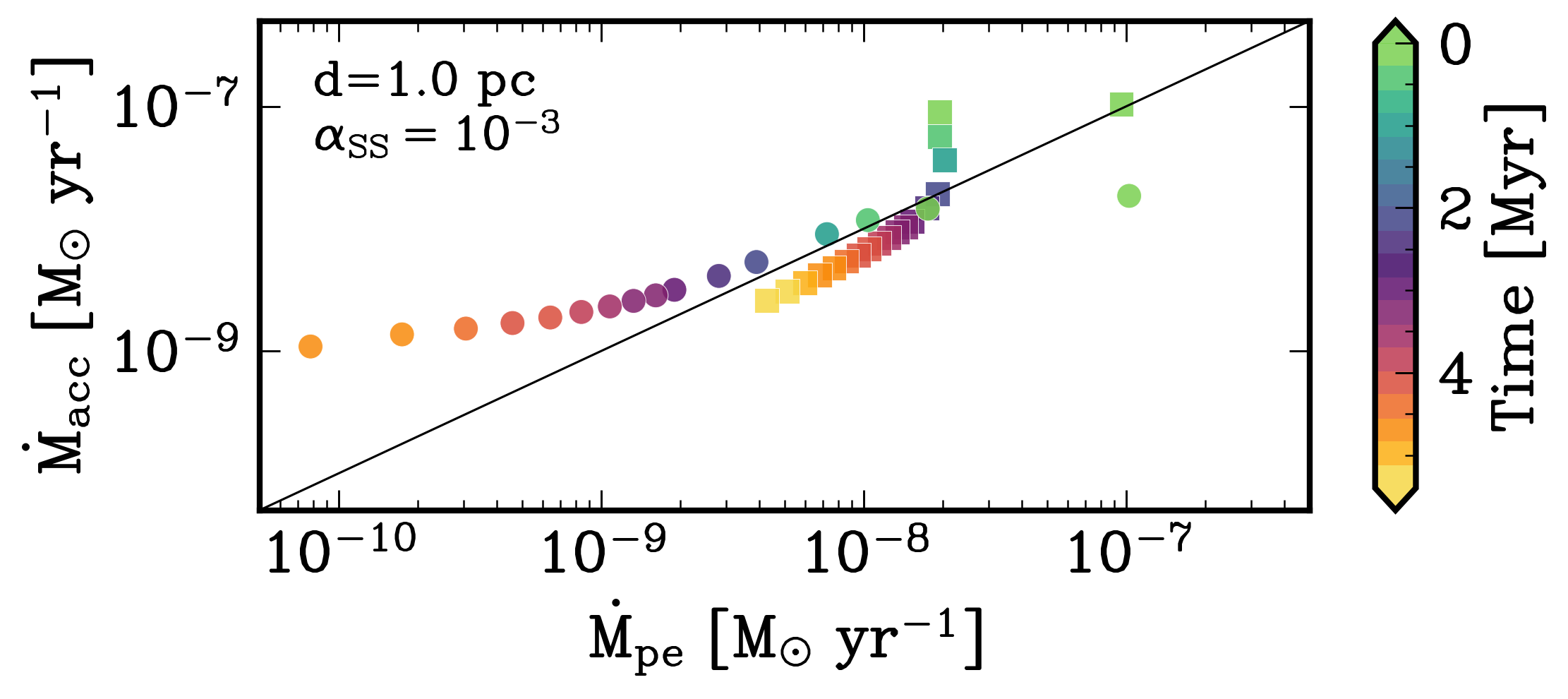}
    \caption{Evolution of the mass accretion and photoevaporative mass-loss rates of the discs shown in Figures~\ref{fig:disc_evolution_0.3pc_alpha1e-3} (top), \ref{fig:disc_evolution_0.3pc_alpha5e-4} (middle) and \ref{fig:disc_evolution_1.0pc_alpha1e-3} (bottom). Squares are used for the viscous evolution, while circles for the MHD wind-driven evolution. The lighter, slightly larger points in the top panel indicate the mass accretion rates of a disc unaffected by external photoevaporation, thus emphasizing the impact of the environment on the evolution of the inner disc, especially in the viscous scenario. In contrast, for a disc evolving  through an MHD wind, these points are indistinguishable from those undergoing external photoevaporation. It is important to note that, in this case, the mass-loss rates hold no physical significance, as the disc is not losing material due to photoevaporation. The black diagonal line in each panel indicates the $\dot{M}_{\rm pe} = \dot{M}_{\rm acc}$ line.}
    \label{fig:macc_mwind}
\end{figure}

As shown in the previous section, external photoevaporation has an important impact in shaping the evolution of protoplanetary discs and their properties. As we will explore in this section, it also has clear and direct implications for the evolution of the mass accretion rate, which differs for viscous and MHD wind-driven discs. 
In Figure~\ref{fig:macc_mwind} we compare the mass accretion rate and the photoevaporative mass-loss rate as a function of time, for both the viscous and wind-driven discs presented in Figures~\ref{fig:disc_evolution_0.3pc_alpha1e-3} (top panel), \ref{fig:disc_evolution_0.3pc_alpha5e-4} (middle panel) and \ref{fig:disc_evolution_1.0pc_alpha1e-3} (bottom panel). The dots are colour-coded by the disc's age, with earlier times shown in green and later times shown in orange/yellow. 
A viscously evolving disc experiences a significant decrease in mass accretion rate, due to external photoevaporation driving faster evolution over its 5+Myr lifetime. This is illustrated in the upper panel of Figure~\ref{fig:macc_mwind}, which shows a drop of almost three orders of magnitude in the mass accretion rate. However, the mass accretion rate of the MHD wind-dominated disc is almost constant throughout the evolution, only varying by roughly an order of magnitude over the disc lifetime. This is because the local disc surface density, close to the star, driving accretion, is almost unperturbed, as illustrated by previous results (see the middle panel of Figure~\ref{fig:disc_evolution_0.3pc_alpha1e-3} and Section~\ref{sec:disc_properties}). An important point to emphasise here is also the clear linear correlation between the photoevaporative mass loss rate and the mass accretion rate in the viscous case, which would not exist without external photoevaporation. The squares align closely with the black solid line, which represents the $\dot{M}_{\rm pe} = \dot{M}_{\rm acc}$ relation. 
To further highlight this point, we also plot the values of mass accretion rates for a disc not exposed to external irradiation. This is represented by the lighter points in the top panel of Figure~\ref{fig:macc_mwind}: we note a significant difference for the viscous case and the absence of a one-to-one correlation, while for the MHD wind case the points are indistinguishable from a disc with external irradiation. 
The photoevaporative mass-loss rates span roughly three orders of magnitude in intensity, but they remain comparable between the two scenarios. This reflects the evolution of the disc's outer radius observed in Section~\ref{sec:disc_properties} and Figure~\ref{fig:evolution_rout}, being the photoevaporative mass-loss rate a function of the disc's size. 

A disc with lower viscosity evolves more slowly. As a result, the disc has overall higher mass accretion rates at fixed time compared to a disc with higher turbulent viscosity, as illustrated in the middle panel of Figure~\ref{fig:macc_mwind}. This may seem counter-intuitive, however, the main reason is that a slower viscous evolution (due to low viscosity) leads to a slower evolution of the disc outer edge and therefore the photoevaporative mass-loss rate remains higher. As a consequence, the accretion rate also remains higher to balance the loss at the outer edge, and the same linear correlation between mass-loss and mass accretion rates is observed for the viscously evolving disc.  
The primary distinction between the viscous disc and the MHD wind-dominated disc lies in their mass-loss rates. Since the viscous disc tends to be larger overall, its mass-loss rates are typically higher than those of the MHD wind case. 

Even when the impact of external photoevaporation is lower (for discs at larger distances from the massive star), we find a similar trend to the other cases. The mass accretion rate does not vary as much over the disc lifetime, which is particularly evident for the MHD winds scenario. This is illustrated in the lower panel of Figure~\ref{fig:macc_mwind}. The photoevaporative mass-loss rates, however, span roughly a few orders of magnitude by the time the disc has reached 5~Myr, for the MHD wind case. Although slightly less apparent in this case, we note once again that, for the viscous disc, the mass accretion rate linearly correlates with the photoevaporative mass-loss rate after a few Myr. Irrespective of the different physical properties of the system, a viscously evolving disc will always evolve towards $\dot{M}_{\rm pe} \approx \dot{M}_{\rm acc}$. 

% It is worth noting here that we let the discs evolve for up to 5~Myr. The difference in the evolution of the two discs is more evident at later times, since the disc has had time to reach an "equilibrium" between the mass-loss and mass accretion rates. 
% If we were to compare our results with a younger star forming region, like the ONC (age $\sim 1-3$~Myr), 

\subsection{A test on viscous evolution}
\begin{figure*}
    \centering
    \includegraphics[width=0.85\textwidth,trim=0cm 0cm 0cm 0cm,clip]
    {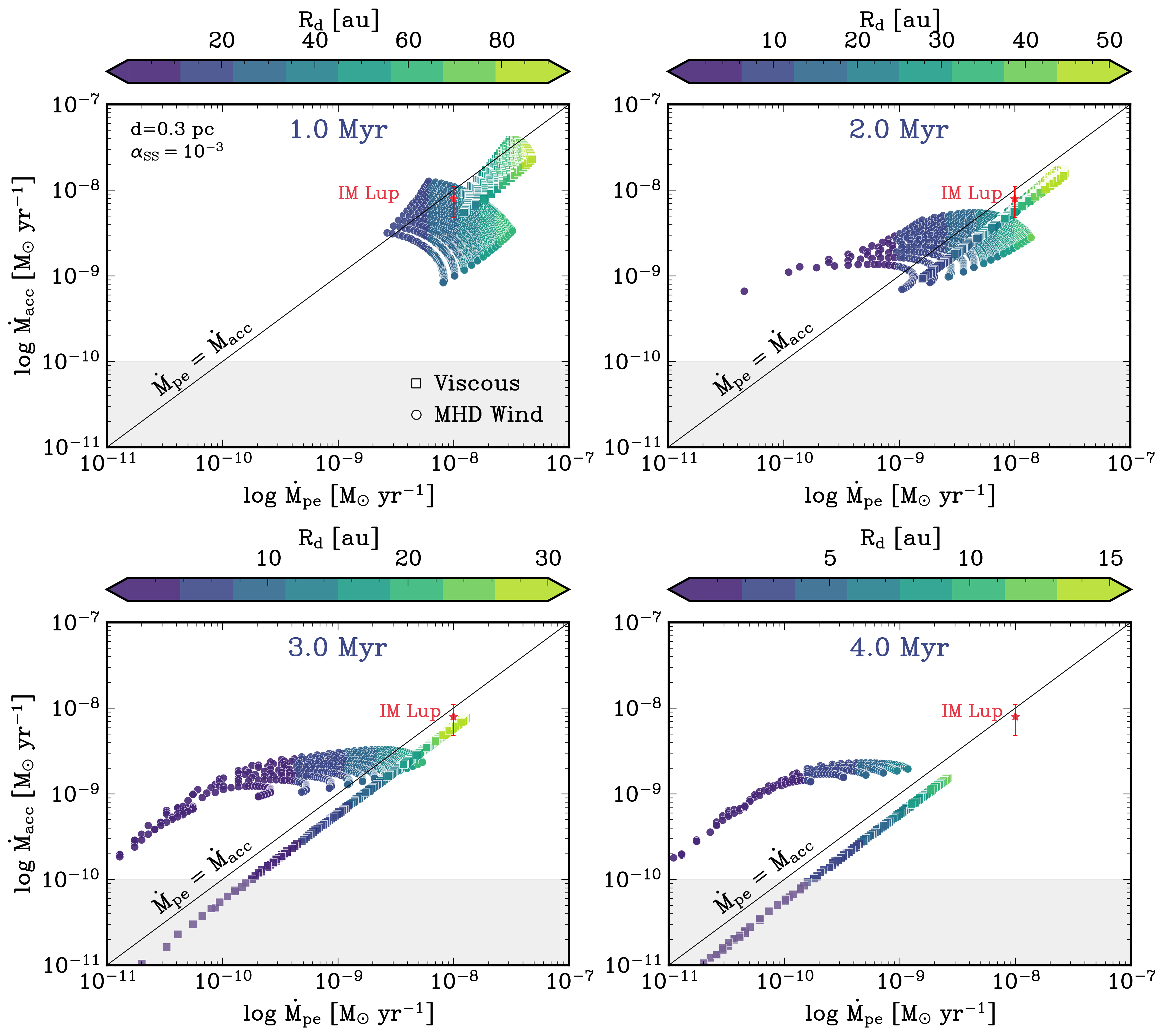}
    \caption{Distribution of mass accretion and photoevaporative mass-loss rates for a sample of discs (with fixed $\alpha_{\rm SS}=10^{-3}$ and $d=0.3$~pc), across different ages (the age is indicated at the top of each panel, in blue). Squares are used for the viscous evolution, while circles for the MHD wind-driven evolution. The points are colour-coded by the disc size at each corresponding time step. 
    The light grey area in each panel represents the threshold below which mass accretion rates are hard to measure \protect\citep[e.g.,][]{Muzerolle2003}. The black line indicates the $\dot{M}_{\rm pe} = \dot{M}_{\rm acc}$ line, while the red star shows the position of IM~Lup on the diagram, with corresponding error bars for the mass accretion rate \protect \citep{2017A&A...600A..20A}.}
    \label{fig:macc_mwind_ndiscs}
\end{figure*}
% \begin{figure}
%     \centering
%     \includegraphics[width=\columnwidth,trim=0cm 0cm 0cm 0cm,clip]{figures/macc_mwind_alpha0.001_psi10000.0_lam3.0_2.5myr.png}
%     \caption{Distribution of mass accretion and mass-loss rates for a sample of discs (with fixed $\alpha_{\rm SS}=10^{-3}$ and $d=0.3$~pc), which have evolved for 2.5~Myr. Squares are used for the viscous evolution, while circles for the MHD wind-driven evolution. The points are colour-coded by disc size at 2.5~Myr. 
%     The light grey area represents the threshold below which mass accretion rates are hard to measure. The black line indicates the $\dot{M}_{\rm pe} = \dot{M}_{\rm acc}$ line, while the red star shows the position of IM~Lup on the diagram, with corresponding error bars for the mass accretion rate \protect \citep{2017A&A...600A..20A}.}
%     \label{fig:macc_mwind_ndiscs}
% \end{figure}

To demonstrate our results from single models carry over to a generic population, 
we run a set of models with different initial disc properties. The initial disc mass is chosen between 0.05 and 0.2~$M_{\odot}$ (in steps of 0.01~$M_{\odot}$), while the initial disc scaling radius is chosen between 10 and 60~au (in steps of 1 au). We obtain a grid of 800 models for a disc around a Solar-like star. The disc's turbulent viscosity is fixed at $\alpha_{\rm SS} = 10^{-3}$ and we also set the disc population at 0.3~pc from the irradiating source and evolve the discs for up to 4.5~Myr. 
% This is the earliest time at which we roughly observe the linear trend developing between mass-loss and mass accretion rates (see Figure~\ref{fig:macc_mwind}. 
Our key results are illustrated in Figure~\ref{fig:macc_mwind_ndiscs}, which shows the distribution of mass accretion and photoevaporative mass loss rates for a sample of discs evolving under pure viscosity or MHD winds, across different ages (1.0, 2.0, 3.0 and 4.0~Myr). Squares represent viscous evolution, while circles indicate MHD wind-driven discs, colour-coded by the disc size at each corresponding age. The black line marks the $\dot{M}_{\rm pe} = \dot{M}_{\rm acc}$ relation, while the grey-shaded region denotes the regime where mass accretion rates become difficult to measure \revision{\citep[e.g.,][]{Muzerolle2003, 2011ApJ...743..105I, 2013A&A...551A.107M, 2024A&A...690A.122C}}. 
Initially (1.0~Myr), most discs lie below the $\dot{M}_{\rm pe} = \dot{M}_{\rm acc}$ line, with larger discs exhibiting higher mass-loss rates. 
Over time, the viscous evolving discs exhibit a sharp decrease in $\dot{M}_{\rm pe}$ and consequently in $\dot{M}_{\rm acc}$, while also contracting significantly. We then observe a tight linear correlation between mass-loss and mass accretion rates. All discs in this population evolve towards the one-to-one line, irrespective of their initial conditions. 
In contrast, the MHD wind-driven discs maintain higher accretion rates and similar values of photoevaporative mass-loss rates in comparison to the viscous discs, while presenting overall smaller sizes. In this case, there is no general expectation that this $\dot{M}_{\rm acc} \sim \dot{M}_{\rm pe}$ correlation will hold, which is consistent with the trends found in our numerical simulations.

\subsubsection{The case of IM~Lup} \label{sec:imlup}
Figure~\ref{fig:macc_mwind_ndiscs} also shows the location in the $\dot{M}_{\rm acc}$-$\dot{M}_{\rm pe}$ space of the externally photoevaporating disc IM~Lup, one of the only sources for which both quantities have been reliably estimated. Interestingly, IM~Lup sits on the one-to-one line within the error bars. The mass accretion rate was determined by \cite{2017A&A...600A..20A} using X-Shooter data. 
%, by measuring the flux in excess relative to the photospheric emission, which comes from the release of accretion energy in the form of continuum emission and spectral lines \citep[e.g.,][]{1998ApJ...492..323G, Hartmann1998book}. 
The photoevaporative mass-loss rate, instead, has been estimated from numerical models by \cite{2017MNRAS.468L.108H}. They simulate a disc undergoing external photoevaporation in a weak environment, estimating the photoevaporative flow based on the incident radiation flux and the thermal structure of the gas. By comparing their models to the observed constraints on the disc size and structure \citep{2016ApJ...832..110C}, they infer a mass-loss rate consistent with the expected influence of external photoevaporation in a low- density environment. 
IM~Lup is not a direct analogue of the ONC-like EUV-irradiated cases explored in this work: it resides in a much weaker irradiation environment and is likely undergoing FUV-dominated external photoevaporation. We include IM~Lup here purely as an illustrative example showing that joint constraints on $\dot{M}_{\rm acc}$ and an externally driven $\dot{M}_{\rm pe}$ estimates can be obtained for a real disc, and we emphasise that the expected one-to-one correlation is a generic result, rather than specific to a single photoevaporation scenario.

\section{Discussion}
\label{discussion}

% Viscosity is responsible for redistributing angular momentum within the disc, therefore connecting the innermost regions with the outer part of the disc. 
% The wind dominated discs sit on the left of the $\dot{M}_{\rm acc}=\dot{M}_{\rm wind}$ line. The viscously evolving discs sit on the right! 

Our current study expands on the work of \cite{2020MNRAS.491..903W} by introducing an analytical derivation to explicitly link mass accretion and mass loss rates, as described in Section~\ref{sec:theory_pe}. This analytical approach offers a complementary perspective to the numerical-based results, and allows for a deeper physical understanding beyond the observed trends. Additionally, their study primarily focused on models driven by classical viscous diffusion and did not explore the implications for discs dominated by MHD winds, which we discuss below. Furthermore, we highlight the fact that this experiment between photoevaporation mass-loss rates and accretion rates would be cleaner in an EUV-dominated case, where mass-loss rates are easier to estimate from radio free-free emission. 

Under the effect of external photoevaporation, a viscously evolving disc experiences a more dramatic change in accretion rate over time compared to an MHD wind-driven disc, as shown by the results presented in Figure~\ref{fig:macc_mwind}. This difference can be understood once again in terms of the intrinsic nature of viscous evolution. In a viscous disc, material is accreted onto the central star, while the disc expands outwards due to the redistribution of angular momentum. However, when material is removed from the outer regions by photoevaporative outflows, the inner regions of the disc are inevitably affected. This connection arises because turbulent viscosity within the disc acts to redistribute angular momentum to maintain conservation. As material is accreted from the inner regions of the disc, angular momentum is transported outward and lost in the photoevaporative outflow. Consequently, there is an almost direct relationship between the mass-loss rate due to external photoevaporation and the mass accretion rates onto the host star. 
% This coupling underscores the significant role that external photoevaporation plays in regulating both the mass distribution and accretion of viscously evolving protoplanetary discs. 

The same connection between different regions within a disc is notably absent in MHD wind-dominated evolution. In such discs, in fact, the removal of material from the outer disc edge does not influence the inner regions. This is because angular momentum is primarily extracted directly by the coupling between the gas and the magnetic field, rather than being redistributed internally. As a result, the dynamics of the inner disc remains largely decoupled from the processes occurring at the outer boundary. The removal of angular momentum through MHD winds bypasses the need for viscous transport across the disc, allowing the inner regions to evolve independently of the material lost at the outer edge. This fundamental difference highlights the contrasting mechanisms at play in viscously evolving and MHD wind-driven discs, underlining why mass redistribution and mass accretion behave differently in the two scenarios. 

% Once again, this is can be understood in terms of the nature of a viscously evolving disc. While material is being accreted onto the central star, the disc has to expand due to redistribution of angular momentum. But if material is then removed from the outer regions of the disc, the inner region is also affected. Turbulent viscosity acts to redistribute angular momentum within the disc itself in order to conserve it, and, therefore, the more material is removed from the outer disc, the more is also accreted to counterbalance the loss. The consequence is therefore an almost one-to-one correlation between mass-loss rates due to external winds and mass accretion rates. 

%\gb{remove-> ? It is worth emphasising here that the mass accretion rates in a viscously evolving disc are generally higher than those of a disc dominated by MHD winds, especially at early times. This distinction might lead to insights into the driving mechanism and disc properties, based on current observations. If mass accretion rate measurements were sufficiently accurate and found to be relatively high, it would suggest a viscous evolution for young discs. Furthermore, as shown in Figure~\ref{fig:macc_mwind}, viscously evolving discs exhibit a significant difference in mass accretion rates at later times. With precise measurements of disc ages, mass-loss rates, and mass accretion rates at a fixed distance, it would be possible to place constraints on disc viscosity.} 

Figure~\ref{fig:macc_mwind_ndiscs} provides a powerful diagnostic for distinguishing between viscous and MHD wind-driven disc evolution. In the case of viscous discs, mass accretion and photoevaporative mass-loss rates maintain a clear near one-to-one correlation. Very young discs are initially dominated by mass removal by external photoevaporation ($\dot{M}_{\rm pe} > \dot{M}_{\rm acc}$), because the viscous time scale is much longer than the timescale over which the outer edge of the disc evolves. Once $t_{\nu} < t_{\rm edge}$, the loss of material at the outer edge is balanced by the accretion onto the central star, and the viscous discs sit close to the $\dot{M}_{\rm pe} = \dot{M}_{\rm acc}$ line. 
In contrast, MHD wind-driven discs exhibit no such \revision{\st{correlation}} one-to-one correlation. Early in their evolution, external photoevaporation removes mass at a higher rate than accretion onto the star. However, as they evolve, they gradually shift to a region where $\dot{M}_{\rm acc} > \dot{M}_{\rm pe}$, meaning accretion remains efficient while mass-loss at the outer edge becomes less significant. Crucially, this transition occurs over a relatively long timescale, implying that the clear separation between viscous and MHD wind evolution only becomes evident in older discs. 

Given that most of the observed externally irradiated discs, such as those in the ONC, are still relatively young ($\sim 1-3$~ Myr), we could see systems that have not yet reached this late-time regime. This is particularly evident for the larger MHD wind-dominated discs, which sit in a region where $\dot{M}_{\rm acc} < \dot{M}_{\rm pe}$ and exhibit strong photoevaporative mass-loss rates. 
%Consequently, differentiating between viscous and MHD wind-driven evolution may be more difficult than expected. 
Nonetheless, the linear correlation observed for viscous discs appears as early as $\sim 2$~Myr, even earlier in some cases depending on the intensity of turbulence and the distance from the irradiating source. 
For example, in the upper panel of Figure~\ref{fig:macc_mwind}, the tight correlation between $\dot{M}_{\rm pe}$ and $\dot{M}_{\rm acc}$ develops almost as soon as the disc evolves, around $\sim 1$~Myr. Subsequently, the viscous discs evolve along the line, irrespective of their initial conditions. Therefore, this challenge can be addressed by studying either an older population of discs or a larger sample spanning a range of ages. In both cases, if the discs primarily evolve viscously, they should follow the predicted linear correlation. 

The main implication of this result is that it provides a clear observational test for viscous evolution. With current facilities, it is possible to obtain accurate measurements of both mass accretion and photoevaporative mass-loss rates, enabling direct comparison with theoretical predictions. However, it is important to note here that in a young cluster, stars are in constant motion due to gravitational interactions. As a result, proplyds do not remain at a fixed distance from massive stars but instead experience time-dependent UV exposure. This motion was suggested as one of the solutions to the proplyds lifetime problem \citep{2019MNRAS.490.5478W}. Because older stars migrate outward and younger stars inward within a cluster, young stars, which also tend to have a larger reservoir of mass in their discs, are preferentially exposed to external irradiation by massive stars at the centre of the cluster. Consequently, these younger stars contribute more significantly to the survival fraction of protoplanetary discs. This non-uniform, time-dependent exposure means that mass-loss rates fluctuate over time, allowing some discs to survive much longer than predicted. Thus, the evolving spatial distribution of stars within the cluster effectively extends the lifetime of many proplyds beyond the theoretical estimates. This also implies that if stars are on highly radial orbits, their instantaneous position within the cluster may not accurately reflect their time-averaged UV exposure. As a result, measurements of mass-loss rates might not be representative of their long-term averaged evaporation history, and the expected correlation between mass-loss and accretion rates would be erased. 
\revision{Previous studies have also coupled disc evolution to cluster dynamical histories in order to interpret externally driven dispersal \citep[e.g.][]{2014MNRAS.441.2094R, 2017A&A...604A..91W}, and extending such approaches would be a natural next step.} Therefore, in the future coupling disc evolution calculations to dynamical calculations of the cluster for both viscous and MHD-wind dominated cases would be valuable to explore the role of this confounding factor. 

\subsection{Current observations of mass accretion rates and photoevaporative mass-loss rates}
\begin{figure}
    \centering
    \includegraphics[width=0.9\columnwidth,trim=0cm 0cm 0cm 0cm,clip]{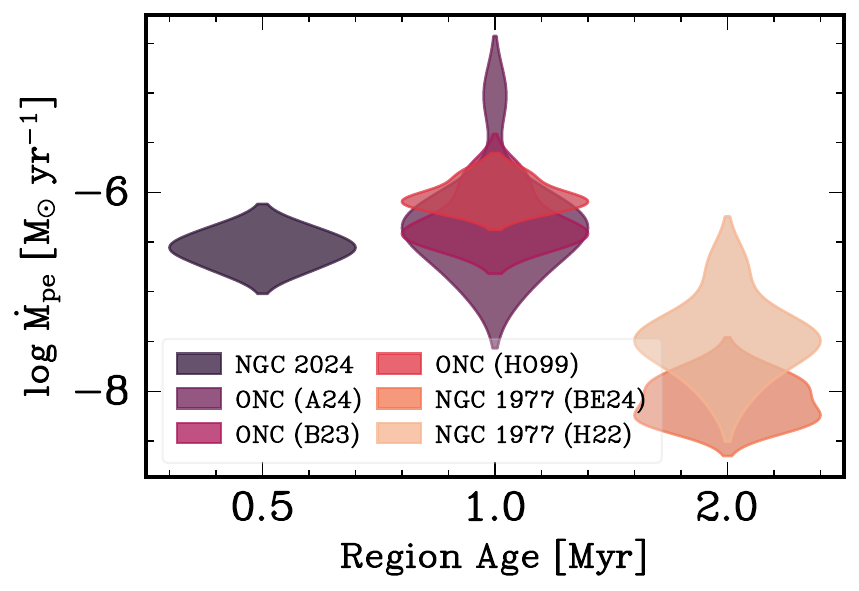}
    \caption{Distribution of photoevaporative mass-loss rates in regions of different ages. Data are taken in order as they appear in the legend from \protect \cite{2021MNRAS.501.3502H, 2024A&A...687A..93A, 2023ApJ...954..127B, 1999AJ....118.2350H, 2024ApJ...967..103B, 2022MNRAS.512.2594H}. Ages are taken from \protect \cite{planetformationenvironmentscollaboration2025}.}
    \label{fig:mwind_age_violin}
\end{figure}
Mass accretion rates are typically derived from spectroscopic observations of emission lines, such as H$\alpha$, which trace the accretion shock onto the stellar surface \revision{\citep{Hartmann1998book, Muzerolle2003, 2017A&A...600A..20A, 2025A&A...704A..42F}}. These lines are used to estimate the accretion luminosity, which is then converted into a mass accretion rate using empirical calibrations \citep{hartmann2016}. 
\cite{2012ApJ...755..154M} conducted a comprehensive study using the Hubble Space Telescope to observe approximately 700 sources within the ONC. They employed photometric data to construct the spectral energy distribution (SED) for each source, enabling the determination of stellar parameters such as mass and age. The mass accretion rate was estimated by analysing the ultraviolet excess emission, which indicates ongoing accretion processes. Their findings revealed that $\dot{M}_{\rm acc}$ increases with stellar mass and decreases over time. 
\revision{We note explicitly that the UV-excess in \citet{2012ApJ...755..154M} is derived from U-band photometry rather than from a flux-calibrated UV spectrum. More recently, \citet{2025A&A...703A.133P} revisited UV-excess measurements in the ONC with improved coverage, and found broadly comparable $\dot{M}_{\rm acc}$ values spanning $\sim 10^{-10}$--$10^{-7}\,M_{\odot}\,\mathrm{yr^{-1}}$ depending on stellar mass and extinction assumptions. This range is useful context for comparison with externally-driven $\dot{M}_{\rm pe}$ values reported for ONC proplyds.}

Mass-loss rates from proplyds, instead, have been determined from observations by analysing emission lines, particularly those from ionised gas (e.g., H$\alpha$, [OIII], [NII]), which trace the photoevaporative flow. By measuring the surface brightness and spatial distribution of these lines, the electron density and velocity of the ionised flow can be estimated. These observational constraints, when combined with theoretical models, yield photoevaporative mass-loss rates estimates in the range of $10^{-8} - 10^{-6} M_{\odot} yr^{-1}$. Figure~\ref{fig:mwind_age_violin} shows the distribution of measurements of photoevaporative mass-loss rates in regions of different ages. 

\cite{1999AJ....118.2350H} utilised high-resolution imaging and spectroscopy to study proplyds in the ONC. They measured the ionized gas emission to estimate photoevaporative mass-loss rates. Their analysis indicates that mass-loss rates due to external photoevaporation range from $10^{-7}$ to $10^{-6} M_{\odot} yr^{-1}$, depending on the distance from the irradiating star and the incident UV flux. More recently, \cite{2024A&A...687A..93A} used VLT/MUSE observations of 12 proplyds in the ONC, some of which had already been studied by \cite{1999AJ....118.2350H}. They measured ionization front radii in H$\alpha$, [OI], [OII], and [OIII] emission lines. Assuming ionization equilibrium, they estimated mass-loss rates by correlating ionization front sizes with incident UV radiation. They found rates broadly consistent with the previous findings of \cite{1999AJ....118.2350H}. 
\revision{Complementary constraints also come from regions spanning a range of irradiation conditions and ages, including $\sigma$~Orionis \citep{2023A&A...679A..82M}, NGC~2024 \citep{2021MNRAS.501.3502H}, and NGC~1977 \citep{2022MNRAS.512.2594H, 2024ApJ...967..103B}.}
\cite{2021MNRAS.501.3502H} estimated mass-loss rates in NGC~2024 using models from the FRIED grid, which simulates FUV driven external photoevaporation under varying UV radiation. By matching observed \revision{ionisation front} sizes and distances to ionising sources, they inferred lower mass-loss rates than in the ONC, reflecting the region's weaker UV flux. In NGC~1977, \cite{2022MNRAS.512.2594H} combined ionised gas measurements with FRIED models to estimate photoevaporative mass-loss\revision{, building on earlier work in similar environments \citep{2016ApJ...826L..15K}}. 

For EUV-driven external photoevaporation specifically, mass-loss rates can be primarily inferred through radio free-free emission and hydrogen recombination lines \citep[e.g.][]{Churchwell1987}. The ionized gas in the outflow produces free-free radiation, which can be detected at millimetre to centimetre wavelengths using facilities like ALMA (Atacama Large Millimetre Array) and VLA (Very Large Array). 
% This emission traces the ionized wind originating from the outer disc, with higher mass-loss rates typically associated with more intense radiation fields. 
% This was investigated by \cite{Pascucci2012} for the nearby disc around TW~Hya, in the case of photoevaporation driven by high-energy photons from the central star. 
\cite{2023ApJ...954..127B} investigated the proplyds in the ONC, using ALMA Band 3 observations. In this work, they spatially resolve dust emission from the discs and free-free emission from the ionization fronts of 12 proplyds. Mass-loss rates were determined using two independent methods. The first assumes ionization equilibrium, where the recombination rate balances the ionization rate from EUV radiation, allowing for an estimate of the electron density. Mass-loss rates are then derived by combining this density with mass continuity equation. The second approach derived the electron density from the brightness of free-free emission. The two methods produce results consistent within an order of magnitude, reinforcing the reliability of these measurements. 

The study by \cite{2024ApJ...967..103B} investigated proplyds in the NGC~1977 star-forming region. Using observations from the Very Large Array (VLA) at multiple radio frequencies (0.3, 6.4 and 15.0~GHz), they identified 34 radio-emitting cluster members, including 10 new candidate proplyds. Photoevaporative mass-loss rates from these proplyds were estimated by analysing the free-free radio emission from the ionized gas in the photoevaporative outflow. The flux density at multiple radio frequencies was measured to identify sources dominated by optically thin free-free emission. Assuming a characteristic electron temperature of $T_e \sim 10^4$~K, they calculated the emission measure and the electron density. The resulting mass-loss rates of $\sim 10^{-9} - 10^{-8} M_{\odot} {\rm yr}^{-1}$ are consistent with theoretical predictions for discs exposed to intermediate UV radiation fields, like those in NGC~1977. 

\revision{Hydrogen recombination-line diagnostics provide a complementary route to constraining the ionised outflow, and recent work has begun to explore their potential for externally irradiated discs beyond the free-free continuum alone \citep[e.g.,][]{2025ApJ...983...81B}.}

\revision{While $\dot{M}_{\rm acc}$ and $\dot{M}_{\rm pe}$ can be each be measured in favourable cases, reliable and simultaneous detections of both mass accretion and mass-loss rates do not exist, with the exception of IM~Lup. The brightest EUV-irradiated proplyds provide the cleanest $\dot{M}_{\rm pe}$ measurements (e.g. from resolved free-free emission), but in these objects standard accretion tracers can be contaminated by emission from the ionisation front. Conversely, in weaker irradiation environments accretion diagnostics are typically cleaner, but the ionised outflow becomes faint and $\dot{M}_{\rm pe}$ is harder to detect directly. This selection mismatch currently limits population-level applications, and motivates coordinated radio and spatially resolved spectroscopic surveys.} Therefore, our study motivates the further measurements to advancing our understanding of disc evolution. In particular, we strongly advocate for mass accretion measurements of proplyds in the ONC and NGC~1977 thought to be undergoing EUV photoevaporation, which already have photoevaporative mass-loss rates determined from radio imaging.

\section{Conclusions} \label{sec:conclusions}
This work explores the impact of EUV-driven external photoevaporation on protoplanetary disc evolution, with findings that also apply to the FUV-dominated case. 
We show that external photoevaporation significantly alters the evolution of a viscous disc, leading to rapid depletion of the disc mass and a shrinking in size. Conversely, MHD wind-driven discs remain relatively unaffected by the external mass-loss, maintaining a nearly constant accretion rate over time. This fundamental distinction between viscous and MHD wind-driven discs leads to significant differences in the correlation of their mass accretion and external photoevaporative mass-loss rates. 

Our numerical simulations, also supported by analytical estimates, confirm that viscous discs show a strong one-to-one correlation between mass accretion rates and photoevaporative mass-loss rates, while MHD wind-dominated discs show no such relationship. Mass accretion rates and photoevaporation rates can be correlated in the MHD-wind scenario, but this is purely an implicit correlation as both decline with age, but we do not find one-to-one cases in our models. 
% This provides a clear observational test for identifying viscously evolving discs, as measuring both accretion and mass-loss rates in various star forming regions can reveal the dominant angular momentum transport mechanism. 
Hence, by comparing the external photoevaporation mass-loss rate with the accretion rate onto the star, we offer a viable way to observationally test in various star forming regions whether protoplanetary discs undergo viscous evolution. Therefore, while our modelling of MHD-wind driven discs is phenomenological, the predicted evolution should not be taken as representative. Our suggested approach can be considered as a null hypothesis test on whether disc evolution is viscous-driven or not. In the case we can reject the null hypothesis that disc's are viscously evolving, an MHD wind-driven scenario is then the likely alternative.  

These results highlight the importance of external photoevaporation as a key factor in shaping disc evolution and provide a promising avenue for future observational studies. Direct measurements of accretion and photoevaporative mass loss rates in the same externally irradiated discs will be crucial to testing these predictions and clarifying our understanding of disc evolution processes. 

\section*{Acknowledgements}
\revision{We thank Carlo Manara for comments and questions that greatly improved the manuscript.}
Additionally, we wish to thank Anna Penzlin, Andrew Winter and Giovanni Rosotti for fruitful discussions and comments that improved this work. 
GB and JEO were funded by the European Research Council (ERC) under the European Union’s Horizon 2020 research and innovation programme (Grant agreement No. 853022, PEVAP). 

%%%%%%%%%%%%%%%%%%%%%%%%%%%%%%%%%%%%%%%%%%%%%%%%%%
\section*{Data Availability}

The data underlying this article will be shared on reasonable request to the corresponding author.



\bibliographystyle{mnras}
\bibliography{references} % if your bibtex file is called example.bib

%\begin{thebibliography}{99}
%\end{thebibliography}

%%%%%%%%%%%%%%%%%%%%%%%%%%%%%%%%%%%%%%%%%%%%%%%%%%

%%%%%%%%%%%%%%%%% APPENDICES %%%%%%%%%%%%%%%%%%%%%

\appendix 

\section{calculation of the coefficients and the final solution to the disc evolution equation}
\label{appx:coefficients_an}
The coefficients $A_n$ can be calculated using the normalisation condition for the total disc mass and assuming an initial condition for the disc surface density \citep{Clarke2001},
\begin{equation}
    \Sigma(x,0) = \frac{M_{\rm d}}{2\pi R_{\rm g}^2} \frac{e^{-x^2}}{x^2}.
\end{equation}
We thus obtain an initial condition for $S(x,\tau)$,
\begin{equation}
    S(x,0) = \sum_n A_n {\rm sin}(k_n x) = \frac{M_{\rm d}}{2\pi R_{\rm g}^2} x e^{-x^2}.
\end{equation}
The coefficients $A_n$ can now be determined \footnote{Using the formula for the Fourier cosine series.}, yielding
\begin{equation}
    \begin{split}
        A_n = & \frac{2}{x_{\rm d}} \int_0^{x_{\rm d}} S(x,0) {\rm sin}(k_nx) dx \\
            = & \frac{M_{\rm d}}{\pi R_{\rm g}^2 x_{\rm d}} \int_0^{x_{\rm d}} x e^{-x^2} {\rm sin}(k_nx) dx.
    \end{split}
\end{equation}
The integral can then be solved numerically. 

\bsp	% typesetting comment
\label{lastpage}
\end{document}